\documentclass[12pt,dvips]{article}

\usepackage{graphicx,epsfig,subfigure} 
\usepackage{dcolumn}      
\usepackage{bm}           
\usepackage{amsmath,amssymb,url,color,colordvi,chicago,psfrag,framed,mathtools}

\newcommand{\tit}[1]{\textit{#1}}   
\newcommand{\vm}[1]{\mathbf{#1}}
\newcommand{\vx}[1]{\vm{x}}
\newcommand{\fref}[1]{Figure~\ref{#1}}
\newcommand{\tref}[1]{Table~\ref{#1}}
\newcommand{\eref}[1]{(\ref{#1})}

\newcommand{\Rv}{\mathbf{R}}
\newcommand{\av}{\mathbf{a}}

\newcommand{\xv}{\mathbf{x}}
\newcommand{\xpv}{\mathbf{x'}}
\newcommand{\tauv}{\boldsymbol{\tau}} 

\newcommand{\del}{\nabla}

\newcommand{\rhop}{\rho^+}
\newcommand{\rhom}{\rho^-}
\newcommand{\rhot}{\tilde\rho}
\newcommand{\rhoit}{\tilde\rho_i}
\newcommand{\rhoIt}{\tilde\rho_I}
\newcommand{\rhoJt}{\tilde\rho_J}
\newcommand{\rhoIm}{\rho_I^-}
\newcommand{\rhoImt}{\tilde\rho_I^-}
\newcommand{\rhopt}{\tilde\rho^+}
\newcommand{\rhomt}{\tilde\rho^-}
\newcommand{\Vp}{V^+}
\newcommand{\Vm}{V^-}

\newcommand{\Vt}{\tilde{V}}
\newcommand{\Vit}{\tilde{V}_i}

\newcommand{\VIt}{\tilde{V}_I}
\newcommand{\vIt}{\tilde{v}_I}
\newcommand{\vIm}{v_I^-}
\newcommand{\vImt}{\tilde{v}_I^-}
\newcommand{\valmt}{\tilde{v}_\alpha^-}
\newcommand{\Valmt}{\tilde{V}_\alpha^-}
\newcommand{\VJt}{\tilde{V}_J}
\newcommand{\Vpt}{\tilde{V}^+}
\newcommand{\Vmt}{\tilde{V}^-}

\begin{document}

\title{Linear scaling solution of the all-electron Coulomb problem in 
solids}

\author{J.~E. Pask$^{1,}$\footnote{Corresponding author. 
E-mail: \texttt{pask1@llnl.gov}}$\,$, N. Sukumar$^2$ and S. E. Mousavi$^2$}
\date{
$\mbox{}^{1}$Condensed Matter and Materials Division\\
Lawrence Livermore National Laboratory, Livermore, CA 94550.\\
$\mbox{}^2$Department of Civil and Environmental Engineering \\
University of California, Davis, CA 95616. \\
\vspace*{0.2in}
\today}

\maketitle

\begin{abstract}
We present a linear scaling formulation for the solution of the 
all-electron Coulomb problem in crystalline solids. The resulting method is 
systematically improvable and well suited to large-scale quantum 
mechanical calculations in which the Coulomb potential and energy 
of a continuous electronic density and singular nuclear density 
are required. Linear scaling is achieved by introducing smooth, 
strictly local neutralizing densities to render nuclear interactions 
strictly local, and solving the remaining neutral Poisson problem for 
the electrons in real space. While the formulation includes singular
nuclear potentials without smearing approximations, the required
Poisson solution is in Sobolev space $H^1$, as
required for convergence in the energy norm. We employ enriched finite elements, with
enrichments from isolated atom solutions, for an efficient solution of
the resulting Poisson problem in the interacting solid. We demonstrate 
the accuracy and convergence of the approach by direct comparison to 
standard Ewald sums for a lattice of point charges, and demonstrate 
the accuracy in all-electron quantum mechanical calculations with an 
application to crystalline diamond.
\end{abstract}

\section{Introduction}\label{sec1}
The evaluation of the electrostatic potential and total energy of
crystalline solids has been an ongoing problem since the earliest days
of solid state physics~\shortcite{Mad18,Ewa21,WigS33,Fuc35,IhmZC79,Wei81}. 
In \tit{ab initio} density-functional~\shortcite{HohK64,KohS65,JonG89}
calculations, the electrostatic (Coulomb) potential is a sum of
nuclear and electronic contributions. In an infinite crystal,
however, each of these terms diverges and the sum is only
conditionally convergent due to the long-range $1/r$ nature of the
Coulomb interaction. Similarly, the total Coulomb energy is a sum of
electron-nucleus, electron-electron, and nucleus-nucleus
contributions, each of which diverges in an infinite crystal but
combine to yield a finite total electrostatic energy per unit
cell. Hence, in the all-electron quantum mechanical problem in solids, 
there are three distinct divergences which must be addressed
simultaneously: (1) the $1/r$ divergence of the electrostatic
potential at the nuclei, (2) the divergence of both potential and
energy lattice sums due to the the long-range $1/r$ nature of the
Coulomb interaction, and (3) the infinite self energies of the nuclei.

It has been appreciated for some time that the divergences and
conditional convergence of such extended lattice summations can be 
eliminated by formulating the summations in terms of neutral 
densities that are well localized in real and/or reciprocal (Fourier)
space~\shortcite{Ewa21}. In the pseudopotential
approximation~\cite{Pic89}, the nuclei and core electrons are replaced 
by smooth ionic cores, thus eliminating $1/r$ divergences and infinite 
self energies. In the conventional reciprocal space approach for such 
calculations in crystals~\shortcite{IhmZC79,Pic89}, the remaining
lattice sum divergences are eliminated by adding 
neutralizing densities to otherwise divergent Coulomb terms in such a
way that the effects of the added densities cancel in the final expressions. 
Remaining long-range interactions are then rendered short-ranged by 
transforming to reciprocal space, where smooth periodic functions, of 
infinite extent in real space, are well localized. However, the resulting
expressions for the electrostatic potential and total energy contain structure 
factors and/or Ewald sums, and require at least $O(N \log N)$ operations 
to evaluate, where $N$ is the number of atoms in the unit
cell. Moreover, since the approach relies on Fourier transforms, it is difficult 
to implement efficiently on large-scale parallel computational 
architectures due to the need for extensive interprocessor
communications. The limitations of the reciprocal space approach have
inspired much research on real-space and local-orbital based 
approaches~\shortcite{Ari99,Bec00,SolAG02,PasS05b,TorEE06},
which allow for better scaling, a variety of boundary conditions, and 
eliminate the need for Fourier transforms. These approaches accomplish
a linear scaling solution of the Coulomb problem by solving Poisson's
equation in real space, or evaluating the 
associated integral, thus alleviating the need for Fourier transforms, 
and allowing the use of efficient multi-level preconditioning~\cite{Bra77,Bec00}.

In the all-electron quantum mechanical context, however, the
divergences of the nuclear potentials and self-energies must be
confronted in addition to the divergences in lattice sums. Moreover,
the rapid, local variation of the electronic density and potential in
the vicinity of the nuclear singularities must be addressed. One
approach to dealing with nuclear singularities is to approximate the 
singular nuclear densities by finite, localized functions, 
e.g., Gaussians or step functions~\shortcite{MerIB95,ModZK97,GoeI98,WanB00,phanish:2010:NPF}.
This makes possible the solution of a nonsingular total electrostatic
potential, due to both nuclei and electrons, in a single linearly scaling step via solution of Poisson's equation with nonsingular total (nuclear + electronic) density as source term. Furthermore, it makes possible the 
direct evaluation of the total Coulomb energy, with finite
nuclear self-energy that is readily removed. However, as the widths 
of the model nuclear densities are decreased toward physical nuclear 
dimensions (on the order of $10^{-5}$ a.u.) to achieve convergence, the 
potential in the vicinity of the nuclei and nuclear self-energies
become correspondingly large, causing greater absolute errors, which hinder the 
computation of accurate energy differences for different configurations. 
Moreover, finer resolution, and correspondingly more degrees of
freedom, are required to resolve the more rapidly varying potential in the 
vicinity of the nuclei, thus increasing computational cost.  This was 
well demonstrated in a remarkable calculation of the Coulomb 
potential of a uranium dimer~\cite{GoeI98} using a second 
generation interpolating wavelet basis, wherein 22 levels of 
refinement were required to reduce the maximum error in the 
potential to order $10^{-2}$ a.u. These difficulties are a consequence 
of the divergent limit of the sequence: since the three-dimensional 
Dirac-delta is not in $H^{-1}$, the corresponding solution of
Poisson's equation is not in $H^1$, where $H^m$ is the Sobolev space of order $m$.
Therefore, the Coulomb energy diverges 
and a convergent approximation in $H^1$ cannot be constructed.

Other approaches for all-electron quantum mechanical calculations
include singular nuclear contributions analytically and compute
the remaining contributions analytically or numerically, depending on the 
choice of basis. 
One such approach is to compute the electronic contributions 
via solution of Poisson's equation with nonsingular, though rapidly 
varying, electronic density as source 
term~(\shortciteNP{WhiWT89}; \shortciteNP{MurSC92}; 
Tsuchida and Tsukada \citeyearNP{TsuT95};
\citeNP{Bat00}; \citeNP{Bec00}; 
\citeNP{YamH05}; \shortciteNP{TorEE06}; \shortciteNP{BylHW09}; \shortciteNP{LehHP09});
where in crystalline calculations~(Tsuchida and Tsukada \citeyearNP{TsuT95};
\shortciteNP{Bec00,TorEE06}), the density 
must be neutralized to avoid divergent lattice sums. With mesh 
refinement~(\shortciteNP{MurSC92}; Tsuchida and Tsukada \citeyearNP{TsuT95};
\shortciteNP{Bat00,TorEE06,BylHW09,LehHP09})
and/or addition of well-chosen localized functions to the 
basis~\shortcite{YamH05}, one can then solve the resulting 
Poisson equation for the electronic contribution to the all-electron 
potential in a single $O(N)$ scaling step. Integrals involving the 
singular total (electronic + nuclear) potential can be efficiently 
computed using a transformation of the singular nuclear 
part~\shortcite{WhiWT89,MurSC92,Bat00,YamH05}. However, in the
crystalline case \shortcite{TsuT95}, the remaining 
nuclear contribution must then be computed by lattice summation, 
which scales as $O(N^2)$ or $O(N \log N)$ at best. An alternative approach 
to the calculation of the electronic contribution is direct evaluation 
of the associated Coulomb law 
integral~(\shortciteNP{GenDN06}; Juselius and Sundholm \citeyearNP{JusS07}; 
\shortciteNP{WatH08,LosSJ10}). This approach, recently
extended to periodic calculations~\shortcite{LosSJ10}, can 
accommodate a variety of boundary conditions, can attain high 
accuracy, and when combined with the fast multipole 
method~(Greengard and Rokhlin \citeyearNP{GreR87}; \shortciteNP{strain:1996:ALS})
for far-field contributions, can 
achieve $O(N)$ scaling~\shortcite{WatH08}. However, the approach 
can be sensitive to the approximation employed for the singular
$1/|\xv-\xpv|$ kernel~(\citeNP{JusS07}; Watson and Hirao \citeyearNP{WatH08};
\shortciteNP{LosSJ10}), and as with
all such approaches computing just the electronic contribution, it
leaves the singular nuclear contribution to be computed by other means.

Another approach, employed in accurate all-electron density-functional 
electronic structure methods \shortcite{SinN06}, employs a dual 
representation of the density and potential~\shortcite{Rud69,Wei81,Blo94,NikD02}. 
In this approach, the unit cell is partitioned into atom-centered
sphere and interstitial (between spheres) regions. Inside the spheres, where the 
potential is singular and most rapidly varying, a radial-angular
spherical harmonic representation is employed. Outside the spheres, where the
potential is generally smooth, a Fourier representation is used. The
all-electron potential is then computed in two steps: (1) compute the smooth
interstitial potential by standard Fourier techniques, and
(2) solve the boundary value problem in each sphere using Green's functions, 
with boundary values from the previously computed interstitial potential. 
By virtue of the dual representation, this method can provide highly accurate and 
efficient solutions. However, due to the reliance on Fourier
expansions, the scaling is at best $O(N \log N)$ and efficient
parallel implementation is difficult due to the need for extensive
interprocessor communications.

Fast multipole methods~\shortcite{GreR87,strain:1996:ALS} provide an $O(N)$ scaling
solution of the Coulomb problem and can accommodate 
periodic~(Challacombe et al.~\citeyearNP{ChaWH97}; \shortciteNP{KudS98a,KudS00,KudS04})
as well as non-periodic boundary conditions.
In the context of a Gaussian representation of the charge density,
these have become the method of choice for large-scale calculations, as they can 
provide high accuracy, with linear scaling and a moderate 
prefactor, and are well suited to parallel implementation. 
However, the computational cost increases rapidly for 
higher-quality basis sets~\shortcite{KurNH07}, and due to 
the reliance on a Gaussian representation, it is not a general 
purpose method~\shortcite{GenDN06}.

In this paper, we present a systematically improvable, linear scaling
formulation for the solution of the all-electron Coulomb problem 
in crystalline solids. Linear scaling is achieved by 
introducing smooth, strictly local neutralizing densities 
to render nuclear interactions strictly local, and solving 
the remaining neutral Poisson problem for the 
electrons in real space. In so doing, the all-electron 
problem is decomposed into analytic strictly-local 
nuclear, and numerical long-range electronic 
parts; with required numerical solution in $H^1$ so 
that convergence is assured and approximation is optimal 
in the relevant energy norm. Rapid variations in the 
required neutral electronic potential in the vicinity 
of the nuclei are efficiently treated by an enriched 
finite element Poisson solution, using local radial solutions 
as enrichments, thus allowing a $O(N)$ 
scaling solution. The formulation is presented in~Section~\ref{sec2} 
and the solution is elaborated in~Section~\ref{sec3}.
Expressions for the Coulomb energy per unit cell, analytically
excluding the divergent nuclear self-energy, are derived.
In~Section~\ref{sec4}, Coulomb potential and energy calculations for 
two canonical test cases using cubic finite elements (FEs) and enriched 
finite elements (EFEs) are presented. We demonstrate the accuracy 
and convergence of the approach by direct comparison to standard Ewald
sums for a lattice of point charges, and demonstrate the accuracy
in quantum mechanical calculations with an application to crystalline 
diamond. In~Section~\ref{sec5}, the main findings are summarized
and the outlook of the proposed formulation in density-functional 
calculations is indicated.

\section{Formulation}
\label{sec2}
\begin{figure}
\centering
\epsfig{file = 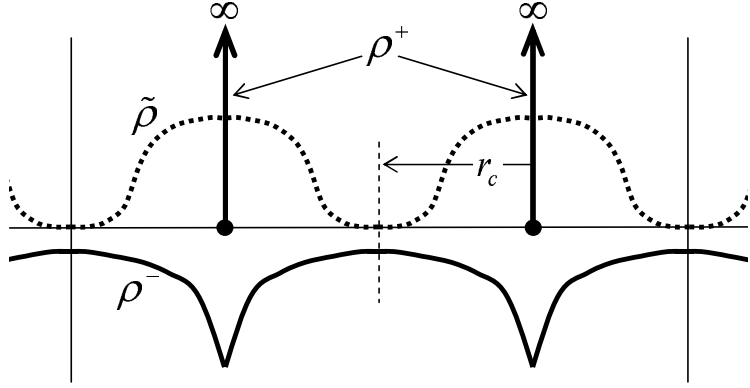, width = 0.6\textwidth}
\caption{Schematic charge density in unit cell. Total density $\rho = \rhop + \rhom$, 
the sum of nuclear point charge and continuous electronic densities. Smooth 
neutralizing density $\rhot$ is introduced to render $\rhop$
short-ranged and $\rhom$ neutral and amenable to direct Poisson
solution. $\rhot$ is constructed as a sum of smooth, localized
densities $\rhot_i$, strictly local within $r=r_c$.}
\label{rhosfig}
\end{figure}

The total Coulomb potential $V(\xv)$ 
diverges at the nuclear positions due to the divergence of the nuclear 
potentials $V_i(\xv)=\frac{q_i}{|\xv-\tauv_i|}$ of charges $q_i$ at 
positions $\tauv_i$ in the unit cell (where atomic units are used here and throughout). 
Moreover, the potentials $\Vp(\xv)$ due 
to all nuclei and $\Vm(\xv)$ due to all electrons in the crystal 
diverge at all points in the cell due to the long-range $1/r$ nature 
of the Coulomb interaction. To resolve these divergences, we introduce 
a smooth neutralizing density $\rhot$ and write the total charge
$\rho$ in the unit cell as
\begin{subequations}\label{eq1}
\begin{align}
\label{eq1a}
\rho(\xv) &= \rhop(\xv) + \rhom(\xv) \\
\label{eq1b}
          &= \underset{\substack{\mathrm{neutralized \ nuclear} \\
                                 \mathrm{charge \ density}}}
                                {\underbrace{\rhop(\xv) - \rhot(\xv)}} 
             + \underset{\substack{\mathrm{neutralized \ electronic} \\
                                   \mathrm{charge \ density}}}
                                  {\underbrace{\rhom(\xv) + \rhot(\xv)}} \\
\label{eq1c}
          & = \rhopt(\xv) + \rhomt(\xv), 
\end{align}
\end{subequations}
where $\rhop(\xv)=\sum_i{\rho_i(\xv)}=\sum_i{q_i \delta(\xv-\tauv_i)}$
is the total nuclear charge density, 
the sum of nuclear densities in the unit cell,
$\rhom(\xv)$ is the electronic charge 
density, and
$\rhopt(\xv)=\rhop(\xv)-\rhot(\xv)$ and $\rhomt(\xv)=\rhom(\xv)+\rhot(\xv)$ 
are the neutralized nuclear and electronic densities, respectively. The
integral over the unit cell $\int_\Omega{\rhot=\sum_i{q_i}}=Q$, 
the total nuclear charge in the cell (Fig.~\ref{rhosfig}).
In order to produce a linear scaling formulation, we form $\rhot$ 
in the unit cell as a sum of smooth, strictly local densities $\rhoIt$ 
centered at atomic positions $\tauv_I$ with 
integrals~$\int{\rhoIt}=q_I$: $\rhot=\sum_I{\rhoIt}$, where the sum 
is over all sites $I$ in the crystal such that 
$\rhoIt \not\equiv 0$ (i.e.~$\rhoIt$ is nonvanishing) in 
the unit cell. 
Since the $\rhoIt$ are strictly localized within a given 
radius $r=r_c$ of each site $I$, i.e., $\rhoIt(\xv)=0$ for $|\xv-\tauv_I|>r_c$, 
the number of terms in the sum varies linearly with the number of 
atoms the unit cell.

The total potential $V(\xv)$ in the unit cell may now be written as
\begin{equation}
V(\xv)=\Vp(\xv)+\Vm(\xv)=\Vpt(\xv)+\Vmt(\xv),
\label{eq2}
\end{equation}
the sum of potentials corresponding to neutralized nuclear and electronic 
densities $\rhopt(\xv)$ and $\rhomt(\xv)$. Let $V_I(\xv)=\frac{q_I}{|\xv-\tauv_I|}$ 
and $\VIt(\xv)$ be the potentials corresponding to charge densities 
$\rho_I(\xv)=q_I\delta(\xv-\tauv_I)$ and $\rhoIt(\xv)$ at each 
site $I$ and let $\rhoIt(\xv)$ be spherically symmetric. Then 
both potential $V_I-\VIt$ and corresponding neutralized density 
$\rho_I-\rhoIt$ vanish beyond $r=r_c$ about site $I$,
since $\VIt = q_I/r$ for $r \ge r_c$ due to the compact-support
of the associated spherically symmetric $\rhoIt(\xv)$.
The total potential in the unit cell associated with neutralized nuclear 
density $\rhopt$ may then be computed as a short-ranged sum in real space:
\begin{equation}
\Vpt(\xv)=\sum_I{V_I(\xv)-\VIt(\xv)}=\sum_I{\frac{q_I}{|\xv-\tauv_I|}-\VIt(\xv)},
\label{eq3}
\end{equation}
where the sum is over all sites $I$ in the crystal such that 
$V_I(\xv)-\VIt(\xv) \not\equiv 0$
in the unit cell. Due to the 
strict locality of $V_I(\xv)-\VIt(\xv)$ around site $I$, 
the number of nonzero terms in the sum varies linearly with 
number of atoms in the unit cell. 

The potential associated with neutralized electronic density $\rhomt$ 
can be obtained from a solution of Poisson's equation:
\begin{equation}
\del^2 \Vmt(\xv) = - 4 \pi \rhomt(\xv),
\label{eq4}
\end{equation}
subject to periodic boundary conditions, with continuous neutralized electronic 
density $\rhomt(\xv)$ as source term. At this point, we note that 
by virtue of the decomposition \eqref{eq1}, the source term $\rhomt$ 
in \eqref{eq4}, unlike the total charge density $\rho$, is in
$H^{-1}$ (indeed, it is $C^0$); so that the corresponding solution $\Vmt$ is in $H^1$, thus 
allowing convergent and optimal approximation in the energy 
norm~\cite{StrF73}. Moreover, the solution can be accomplished in 
$O(N)$ operations in real space~\cite{Bec00}, where $N$ is the number
of atoms in the unit cell. The total all-electron Coulomb potential 
$V(\xv)$ can thus be obtained in $O(N)$ operations using
decompositions (\ref{eq1}) and (\ref{eq2}), without distributed
nucleus or other approximations.

The total Coulomb energy per unit cell in the all-electron case is 
divergent due to the divergence of the nuclear self energies. Thus,
the desired total Coulomb energy excluding nuclear self-energy cannot 
be computed as in the pseudopotential case~\cite{PasS05a} by first 
computing the total Coulomb energy and then subtracting the self-energy. 
Instead, the divergent nuclear self-energy must be excluded analytically. 

\begin{figure}
\centering
\epsfig{file = 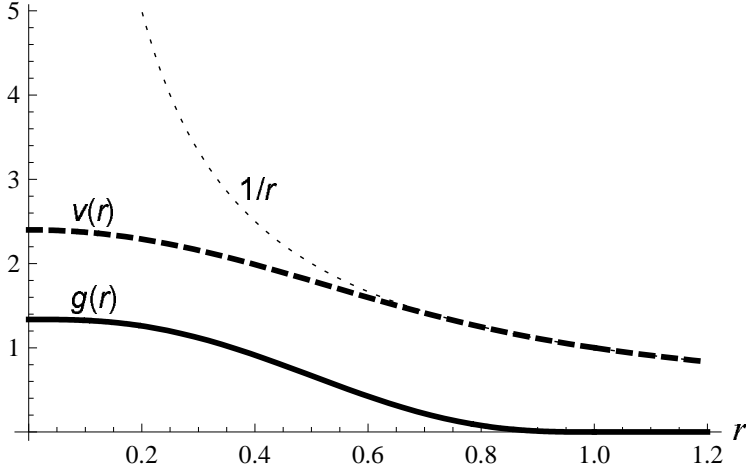, width = 0.6\textwidth}
\caption{Smooth, strictly local unit charge $g(r)$, corresponding potential $v(r)$, and $1/r$, with cutoff radius $r_c=1$. $g(r)=0$ and $v(r)=1/r$ for $r>r_c$.}
\label{gvfig}
\end{figure}

The total Coulomb energy per unit cell can be expressed in terms of 
densities and associated potentials as
\begin{subequations}
\begin{align}
E&=\tfrac{1}{2}\int_\Omega{ d^3x\, \rho(\xv)V(\xv) } \\
\label{eq8}
   &=\tfrac{1}{2}\int_\Omega{ d^3x\, 
      \bigl(\rhopt(\xv)+\rhomt(\xv)\bigr)\bigl(\Vpt(\xv)+\Vmt(\xv)\bigr) } \\
\label{eq9}
   &=\tfrac{1}{2}\int_\Omega{ d^3x\ 
      \bigl(\rhopt(\xv)\Vpt(\xv)+2\rhopt(\xv)\Vmt(\xv)+\rhomt(\xv)\Vmt(\xv)\bigr) } \\
\label{eq10}
   &=E^{++}+E^{+-}+E^{--},
\end{align}
\end{subequations}
where $E^{++}$, $E^{+-}$, and $E^{--}$ correspond to the 
$\rhopt\Vpt$, $\rhopt\Vmt$, and $\rhomt\Vmt$ integrals, respectively, 
and the $\rhopt\Vmt$ interaction term has been retained in favor of 
the equivalent $\rhomt\Vpt$ term to facilitate subsequent integration. 
The divergent self-energy is contained in the $E^{++}$ term. This may 
be extracted as follows:
\begin{equation}
E^{++}=\tfrac{1}{2}\int_\Omega{ d^3x\, \rhopt(\xv)\Vpt(\xv) }
      =\tfrac{1}{2}\int_\Omega{ d^3x\, 
         \sum_I{\bigl(\rho_I(\xv)-\rhoIt(\xv)\bigr)}
         \sum_J{\bigl(V_J(\xv)-\VJt(\xv)\bigr)} },
\label{eq11}
\end{equation}
where the sums are over all atomic positions in the crystal with localized 
densities and potentials $\rhoIt$ and $V_J-\VJt$ nonzero in the unit cell. 
Now, we restrict the neutralizing densities $\rhoIt$ to be
nonoverlapping, i.e., $\rhoIt(\xv)\rhoJt(\xv)=0$ for $I \neq J$. Then
the double summation (\ref{eq11}) reduces to
\begin{align}
E^{++}&=\tfrac{1}{2}\int_\Omega{ d^3x\, 
         \sum_I{\bigl(\rho_I(\xv)-\rhoIt(\xv)\bigr)}\bigl(V_I(\xv)-\VIt(\xv)\bigr) },
    \nonumber \\
      &=\tfrac{1}{2}\sum_i{
         \int{ d^3x\, \bigl(\rho_i(\xv)-\rhoit(\xv)\bigr)}\bigl(V_i(\xv)-\Vit(\xv)\bigr)},
\label{eq12}
\end{align}
the sum of neutralized nuclear self energies, where the last summation
is over atoms in the unit cell and the integral is over all space. Extracting 
the divergent self-energy from (\ref{eq12}), we have 
\begin{align}
E^{++}&=E_\mathit{self}-\tfrac{1}{2}\sum_i{\int{ d^3x\, 
        [\rho_i(\xv)\Vit(\xv)+\rhoit(\xv)\bigl(V_i(\xv)-\Vit(\xv)\bigr)] } },
    \nonumber \\
      &=E_\mathit{self}-\tfrac{1}{2}\sum_i{
        [q_i\Vit(\tauv_i)+\int{ d^3x\, \rhoit(\xv)\bigl(V_i(\xv)-\Vit(\xv)\bigr) }] }.
\label{eq13}
\end{align}
This can be simplified further by employing a common spherically symmetric 
neutralizing charge $\rhoit$ and corresponding potential $\Vit$ at each site. 
Let $\rhoit(\xv)=q_i g(|\xv-\tauv_i|)$ and $\Vit(\xv)=q_i v(|\xv-\tauv_i|)$, 
where $g(r)$ is a smooth radial function strictly localized within $r=r_c$, 
with $\int{g(|\xv|)}=1$ and $r_c$ such that $\rhoit(\xv)$ are nonoverlapping, 
and $v(r)$ is the corresponding potential, as in Fig.~\ref{gvfig}. For $r>r_c$ 
then, $g(r)=0$ and $v(r)=1/r$. Equation~(\ref{eq13}) then becomes
\begin{subequations}
\begin{align}
\label{eq14}
E^{++}&=E_\mathit{self}-\tfrac{1}{2}\sum_i{
        [q_i^2 v(0)+q_i^2\int_0^{r_c}{ dr\, 4\pi r^2 g(r)\bigl(1/r-v(r)\bigr) } ] } \\
\label{eq15}
      &=E_\mathit{self}-\tfrac{1}{2}\sum_i{ (q_i^2 v(0)+q_i^2 I_g) },
\end{align}
\end{subequations}
where $I_g$ is the constant defined by (\ref{eq14}), which depends only on the 
choice of the radial function $g(r)$. In the present work, we employ a
second-derivative continuous $\bigl(g\in C^2(\mathbb{R}_+)\bigr)$ 
piecewise polynomial for $g(r)$:
\begin{subequations}
\begin{align}
\label{eq16}
g(r)&=
   \begin{cases}
   -21 (r - r_c)^3 (6 r^2 + 3 r r_c + r_c^2)/(5 \pi r_c^8),   &   0 \leq r \leq r_c, \\
   \quad 0,   &   r > r_c,
   \end{cases} \\
\intertext{for which}
\label{eq17}
v(r) &=
   \begin{cases}
   (9 r^7 - 30 r^6 r_c + 28 r^5 r_c^2 - 14 r^2 r_c^5 + 12 r_c^7)/(5 r_c^8),   &   0 \leq r \leq r_c, \\
   1/r,   &   r > r_c,
   \end{cases} \\
\intertext{and}
I_g &= 10976/(17875\, r_c).
\label{eq18}
\end{align}
\end{subequations}

From~(\ref{eq9}) and~(\ref{eq10}), the $E^{+-}$ interaction term is given by
\begin{align}
E^{+-}&=\int_\Omega{ d^3x\, \rhopt(\xv)\Vmt(\xv) } \nonumber \\
      &=\int_\Omega{ d^3x\, \sum_I{\bigl(\rho_I(\xv)-\rhoIt(\xv)\bigr)\Vmt(\xv)} } \nonumber \\
      &=\sum_i{q_i \Vmt(\tauv_i)} - \int_\Omega{ d^3x\, \rhot(\xv)\Vmt(\xv) },
\label{eq19}      
\end{align}
where the sum on $I$ extends over all atoms in the crystal with
$\rhoIt \not\equiv 0$
in the unit cell and the sum on $i$ is over atoms in
the unit cell.

Being free of nuclear singularities, the $E^{--}$ term can be
evaluated straightforwardly as
\begin{equation}
E^{--}=\tfrac{1}{2}\int_\Omega{ d^3x\, \rhomt(\xv)\Vmt(\xv) }.
\label{eq20}
\end{equation}
Collecting $E^{++}$, $E^{+-}$, and $E^{--}$ from~(\ref{eq15}),~(\ref{eq19}), 
and~(\ref{eq20}) above, we arrive at the following expression for the
all-electron Coulomb energy per unit cell, excluding nuclear self-energy:
\begin{equation}
E-E_\mathit{self}=\sum_i{ [q_i \Vmt(\tauv_i) - \tfrac{1}{2}q_i^2(v(0)+I_g)] } 
                 + \int_\Omega{d^3x\, (\tfrac{1}{2}\rhomt(\xv) - \rhot(\xv))\Vmt(\xv) },
\label{eq21}                 
\end{equation}
where the integral is over the unit cell and the sum is over atomic 
positions in the cell. Since the densities and potentials in~(\ref{eq21}) 
can be obtained in $O(N)$ operations, as described above, the energy
too, as formulated in (\ref{eq21}), can be obtained in $O(N)$ operations.

In the above expression, the $\rhopt\Vmt$ interaction term was
retained in favor of the equivalent $\rhomt\Vpt$ term to facilitate
analytic integration. However, whereas this eliminates a 
three-dimensional numerical integration, it produces a term in the energy
requiring a pointwise evaluation of the Poisson solution $\Vmt$. In
basis-oriented, variational approaches for the Poisson solution,
however, energy integrals of the solution converge more rapidly than
pointwise values~\cite{StrF73}, and so it is of interest to develop an
alternative expression free of pointwise evaluations of $\Vmt$. This
can be accomplished at the cost of an additional numerical integration
by formulating the $E^{+-}$ interaction term based on the
corresponding $\rhomt\Vpt$ integral rather than $\rhopt\Vmt$ integral 
as in (\ref{eq9}).

The interaction term is
\begin{equation}
\label{eq61}
E^{+-}=\int_\Omega{ d^3x\, \rhopt(\xv)\Vmt(\xv) }
      =\int_\Omega{ d^3x\, \rhomt(\xv)\Vpt(\xv) }.
\end{equation}
This may be expressed as
\begin{align}
E^{+-}
&=\int_\Omega{ d^3x\, [\rhom(\xv)+\sum_I{\rhoIt(\xv)}]\sum_J{\bigl(V_J(\xv)-\VJt(\xv)\bigr) } } 
\nonumber \\
&=\sum_I{ \int_\Omega{ d^3x\, \rhom(\xv)\bigl( V_I(\xv)-\VIt(\xv) \bigr) } } +
  \sum_I{ \int_\Omega{ d^3x\, \rhoIt(\xv) \bigl( V_I(\xv)-\VIt(\xv) \bigr) } } \nonumber \\
&=\sum_i{ \int{ d^3x\, \rhom(\xv) \bigl( V_i(\xv)-\Vit(\xv) \bigr) } } +
  \sum_i{ \int{ d^3x\, \rhoit(\xv) \bigl( V_i(\xv)-\Vit(\xv) \bigr) } }
\label{eq64}
\end{align}
for nonoverlapping densities $\rhoIt(\xv)$, where $I$ extends over 
all contributing sites in the crystal; $i$, over positions in the 
unit cell, and the final integrals are over all space. On letting 
$\rhoit(\xv)=q_i g(|\xv-\tauv_i|)$ and $\Vit(\xv)=q_i v(|\xv-\tauv_i|)$, 
where $g(r)$ and $v(r)$ are as in (\ref{eq16}) and (\ref{eq17}), the 
interaction term becomes
\begin{equation}
\label{eq65}
E^{+-}=\sum_i{ q_i \int_{\Omega_i}{d^3x\, \rhom(\xv)\bigl(1/r_i - v(r_i)\bigr) } } 
+ \sum_i{ q_i^2 I_g },
\end{equation}
where $\Omega_i$ is the sphere of radius $r=r_c$ within which 
density $\rhoit(\xv)$ is localized, $r_i=|\xv-\tauv_i|$, and $I_g$ 
is as in (\ref{eq18}). With this formulation of the interaction term, 
the all-electron Coulomb energy per unit cell, excluding nuclear self
energy becomes
\begin{equation}
\label{eq66}                 
E-E_\mathit{self} \! = \!
\sum_i{ \tfrac{1}{2}q_i^2\bigl(I_g-v(0)\bigr) } + 
\sum_i{ q_i \! \int_{\Omega_i}{\! \! \! d^3x\, \rhom(\xv)\bigl(1/r_i - v(r_i)\bigr) } } + 
\tfrac{1}{2} \! \int_\Omega{\! \! \! d^3x\, \rhomt(\xv)\Vmt(\xv) }.
\end{equation}
In the special case of constant $\rhom(\xv)$, as in the Ewald problem,
the integral in the second term of (\ref{eq66}) 
reduces to $I_{sph} = 14 \rhom \pi r_c^2/75$. 
In the more general case, the weak angular dependence
of $\rhom(\xv)$ about each site $i$ allows for efficient evaluation
using Gaussian quadrature in spherical coordinates~\cite{Str71}.

\section{Solution}\label{sec3}
The computation of the all-electron Coulomb potential and energy as 
formulated in~(\ref{eq2}), (\ref{eq21}), and (\ref{eq66}) 
requires the solution of the Poisson equation (\ref{eq4}) for the 
potential $\Vmt$ corresponding to the neutralized electronic 
charge density $\rhomt$, which may be accomplished by a number 
of methods. Here, we employ an enriched finite element (EFE)
method~\shortcite{StrF73,melenk:pufem,babuska:pufem,suku:2009:bloch} 
in order to efficiently resolve the sharp variations in the 
all-electron densities and potentials while retaining systematic 
improvability and $O(N)$ scaling in the solution process. 
Note that some form of multilevel preconditioning is generally required to achieve 
linear scaling of solution time with respect to number of degrees of freedom (DOFs) 
in FE and related methods. While not demonstrated as yet in the context of EFE methods, 
to our knowledge, it is expected that such preconditioning will apply here also. 
EFE methods add fixed enrichment functions to the classical FE basis in order to 
efficiently represent rapid variations in the solution. This leaves a smooth residual 
(difference between exact solution and enrichment) for the remaining FE basis to 
represent, thus allowing a substantially coarser FE mesh. EFE methods can thus 
be understood as FE methods on softer problems, and so obtain classical 
asymptotic convergence rates with respect to mesh size, depending only on the 
completeness of the classical part of the basis (as shown below). Hence, multilevel 
preconditioning may be expected to apply to EFE as for FE. Indeed, consistent with 
this expectation, linear scaling with respect to number of DOFs has been demonstrated
recently in the context of partition-of-unity enrichment methods~\cite{schweitzer:2003:APM,schweitzer:2008:MAG}.

In the classical FE Poisson solution~\cite{PasS05b}, the potential $\Vmt$ 
is expressed as a linear combination of strictly local, piecewise 
polynomial basis functions $\{\phi_i\}$:
\begin{equation}
\Vmt_{FE}(\xv)=\sum_i \phi_i(\xv) a_i^{FE} .
\label{eq31}
\end{equation}
In the enriched FE solution, a set of functions $\{\psi_\alpha\}$ which 
incorporates \tit{a priori} information about the solution is 
added to the classical basis in order to substantially reduce 
the number of basis functions required to attain a given accuracy; 
so that $\Vmt$ is expressed as
\begin{equation}
\Vmt(\xv)=\sum_i \phi_i(\xv) a_i + \sum_\alpha{\psi_\alpha(\xv)b_\alpha} 
\equiv \sum_k{\Phi_k(\xv) c_k},
\label{eq32}
\end{equation}
where $\alpha$ is summed over the atoms and $\{\Phi_k\}=\{\phi_i\} \cup \{\psi_\alpha\}$ is the enriched FE basis, 
the combined set of classical FE and enrichment basis functions. 

In the present case, the desired solution $\Vmt=\Vm+\Vt$ is the potential 
associated with neutralized electronic charge density $\rhomt=\rhom+\rhot$, 
where $\rhot=\sum_I{\rhoIt}$, the sum of smooth, strictly local, 
nonoverlapping neutralizing densities over atomic sites $I$ in the 
crystal such that 
$\rhoIt \not\equiv 0$
in the unit 
cell. In the all-electron case, the 
electronic density $\rhom$ is large in magnitude, rapidly 
varying, and isolated-atom-like in the vicinity of the 
nuclei while much smaller in magnitude and more smoothly varying 
elsewhere. The neutralizing density $\rhot$ is moderate in 
magnitude and smoothly varying throughout the cell. Hence, the 
rapid variations in the desired solution $\Vmt$ associated with 
neutralized electronic density $\rhomt$ are confined to the 
vicinity of the nuclei where the electronic density $\rhom$ is 
large, rapidly varying, and atomic-like. To increase the efficiency 
of the representation, therefore, we might add to the basis the 
potentials $\vImt$ corresponding to the neutralized electronic 
densities $\rhoImt=\rhoIm+\rhoIt$ in the vicinity of each site $I$; 
where the local electronic densities $\rhoIm$ vary like the total 
electronic density $\rhom$ in the vicinity of site $I$ and integrate 
to the appropriate charge $-q_I$. These may be obtained, for example, 
from a partitioning of the self-consistent electronic density or 
from isolated atom densities. 

For each atom $\alpha$ in the unit cell, then, an \tit{enrichment function}
\begin{equation}
\Valmt(\xv)=\sum_\Rv{ \valmt(\xv-\Rv) }
\label{eq33}
\end{equation}
approximating the desired solution $\Vmt$ in the vicinity of atom $\alpha$ 
is constructed, where $\Rv$ denotes lattice translation vectors such 
that
$\valmt(\xv-\Rv) \not \equiv 0$
in the unit cell. Since the enrichment 
is only needed in the vicinity of atom $\alpha$, the $\valmt$ and 
associated summation are short-ranged. 
The \tit{enriched basis 
functions} $\psi_\alpha$ in (\ref{eq32}) are taken, then, as the
periodic 
enrichment functions $\Valmt$:
\begin{equation}
\psi_\alpha(\xv) = \Valmt(\xv).
\label{eq34}
\end{equation}

To solve the Poisson equation subject to periodic boundary conditions, 
it is sufficient that the basis $\{\Phi_k\}$ satisfy
\begin{equation}
\Phi_k(\xv+\Rv)=\Phi_k(\xv)
\label{eq36}
\end{equation}
for all $\xv$ on the boundary~\shortcite{PasKF99,PasS05b}. Classical 
basis functions $\{\phi_i\}$ satisfying this condition are readily 
constructed~\shortcite{PasKF99}. Furthermore, the enriched basis functions
$\{\psi_\alpha\}$ in (\ref{eq34}) are periodic in the unit cell
by construction (Eq.~\eqref{eq33}). Thus the enriched FE basis $\{\Phi_k\}$ 
as a whole satisfies the required condition also~\cite{suku:2009:bloch}.

Having so constructed the enriched basis $\{\Phi_j\}$, satisfying 
the required boundary conditions, the enriched FE 
solution $\Vmt(\xv)=\sum_j{\Phi_j(\xv) c_j}$ of the Poisson 
equation (\ref{eq4}) with neutralized electronic density 
$\rhomt(\xv)$ is then determined by the sparse, symmetric linear system~\cite{PasS05b}
\begin{subequations}
\begin{align}
\label{eq27}
& \quad \sum_j {L_{ij} c_j} = f_i, \\
\intertext{where}
\label{eq38}
L_{ij} &= \int_\Omega  {d^3x\, \del\Phi_i(\xv) \cdot \del\Phi_j(\xv)}, \\ 
\label{eq39}
f_i &= 4\pi \int_\Omega {d^3x\, \Phi_i(\xv)\rhomt(\xv)}.
\end{align}
\end{subequations}

\section{Results}\label{sec4}
We demonstrate the accuracy and convergence of the formulation presented in 
Section~\ref{sec3} on two canonical test cases: the Ewald
energy of a bcc crystal, for which a reference value is available via standard Ewald 
summation~\cite{Ewa21}, and the all-electron Coulomb potential 
and energy of crystalline diamond. In the numerical computations, the 
parallelepiped unit cell is discretized into 
regular $m \times m \times m $ serendipity cubic (32-node) finite elements.
The shape function expressions for serendipity cubic brick elements are provided
in~\shortciteN{zienkiewicz:2005:TFE}. Upon obtaining radial atomic potential solutions 
$v_\alpha^-$ analytically or from a one-dimensional 
solver (exploiting spherical symmetry),
the neutralized atomic potential functions $\valmt$ are tabulated on a
one-dimensional equi-spaced grid for the Ewald problem and on
a logarithmic grid for the diamond problem.
A quintic spline-fit of the tabulated data is formed and the
enrichment function is constructed as in 
(\ref{eq33}), which is then used in the numerical computations.
Numerical integration for the FE computations is carried out using
$5 \times 5 \times 5$ Gauss quadrature, and
higher-order quadrature is employed in EFE computations.
The FE computation on a $m \times m \times m$ mesh has $7m^3$ degrees of freedom,
and the corresponding EFE computation has just two (number of atoms) more.

\subsection{Ewald problem}
For the classical Ewald energy, we consider a reference bcc crystal 
with unit atomic spacing defined by lattice vectors
\begin{align*}
\av_1&=a(1,0,0), \\
\av_2&=a(0,1,0), \\
\av_3&=a(0,0,1),
\end{align*}
with unit point charges $q_i=1$ located at positions (in lattice coordinates)
\begin{align*}
\tauv_1&=(0,0,0), \\
\tauv_2&=(1/2,1/2,1/2)
\end{align*}
and a constant negative background 
$\rhom=-2/a^3$ bohr$^{-3}$, where $a=2/\sqrt{3}$ bohr. The Ewald 
energy computed by standard $N^2$ scaling summation is $-1.5758343085$ Hartree/atom.

To compute the energy using the $O(N)$ formulation (\ref{eq21}) 
or (\ref{eq66}), we take $g(r)$ as in (\ref{eq16}) with $r_c=1/2$ bohr, 
so that the resulting neutralizing densities $\rhoIt(\xv)= g(|\xv-\tauv_I|)$ 
at each site $I$ in the crystal are as smooth as possible without 
overlapping. The total neutralizing density in the cell is then 
$\rhot(\xv) = \sum_I \rhoIt(\xv)$, where the sum over positions $I$ 
in the unit and nearest neighboring cells is sufficient due to 
the short-range of $g(r)$. 
For the sake of clarity, we distinguish between the cut-off radius
$r_c$ used to represent the neutralizing charge $\rhoIt$ and
that used to form the enrichment function corresponding to the neutralized
electronic charge density. We refer to the former
as $r_{cn}$ and to the latter as $r_{ce}$.
Since the electronic density $\rhom$ is 
constant, we take as local electronic densities $\rhoIm(\xv)=-g(|\xv-\tauv_I|)$ 
with $r_{ce}=1$ a.u.\ so that $\sum_I \rhoIm$ approximates the 
constant $\rhom$ in the cell with $I$ running again over positions 
in the unit cell as well as the nearest neighboring cells. The potentials corresponding 
to neutralized electronic densities $\rhoImt=\rhoIm+\rhoIt$ at each 
site $I$ are then $\vImt=\vIm+\vIt$, where $\vIm(\xv)=-v(|\xv-\tauv_I|)$ 
with $r_{ce}=1$ a.u., $\vIt(\xv)=v(|\xv-\tauv_I|)$ with $r_{cn}=1/2$ a.u., 
and $v(r)$ is as in (\ref{eq17}). The enrichment functions $\Valmt$ 
are then formed as in (\ref{eq33}). The required potential $\Vmt$ 
corresponding to neutralized electronic density $\rhomt=\rhom+\rhot$ 
is obtained from the enriched FE solution of the associated Poisson 
equation subject to periodic boundary conditions, with enrichment 
functions $\Valmt$. The total (electronic + nuclear) potential is then obtained from (\ref{eq2}) 
and the energy, from (\ref{eq21}) or (\ref{eq66}).

\begin{figure}
\centering
\psfrag{rho-}[B][l]{~~~~~$\rho^-$}
\psfrag{rho~}[B][l]{~~~~$\tilde{\rho}$}
\psfrag{rho-~}[B][l]{~~~~~~$\tilde{\rho}^-$}
\psfrag{V+~}[B][l]{~~~~~~$\tilde{V}^+$}
\psfrag{V-~}[B][l]{~~~~~~$\tilde{V}^-$}
\psfrag{V (total)}[B][l]{~~~~~~~~~~~~$V$ (total)}
\mbox{
\subfigure[]{\epsfig{file=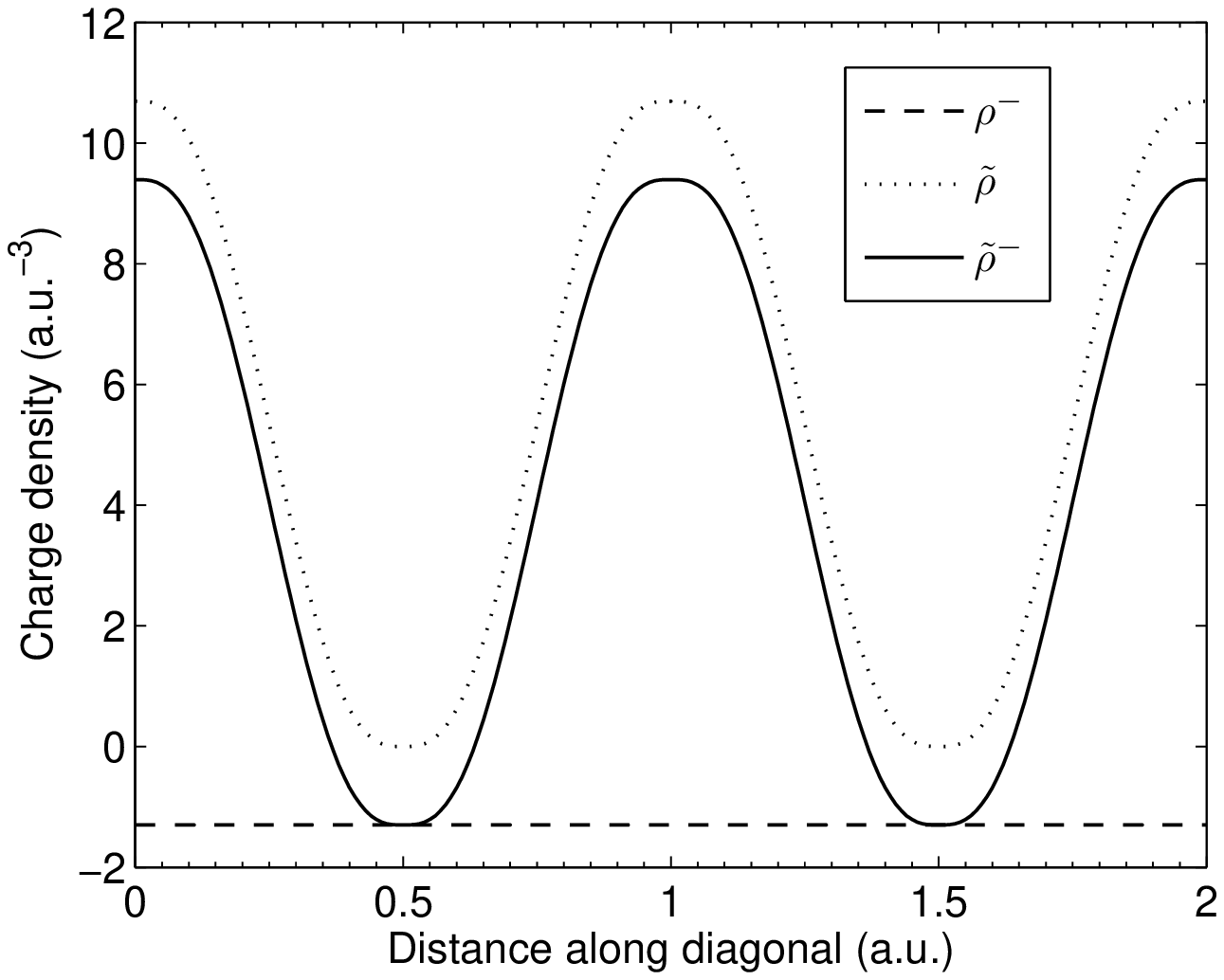, width=0.49\textwidth}}
\subfigure[]{\epsfig{file=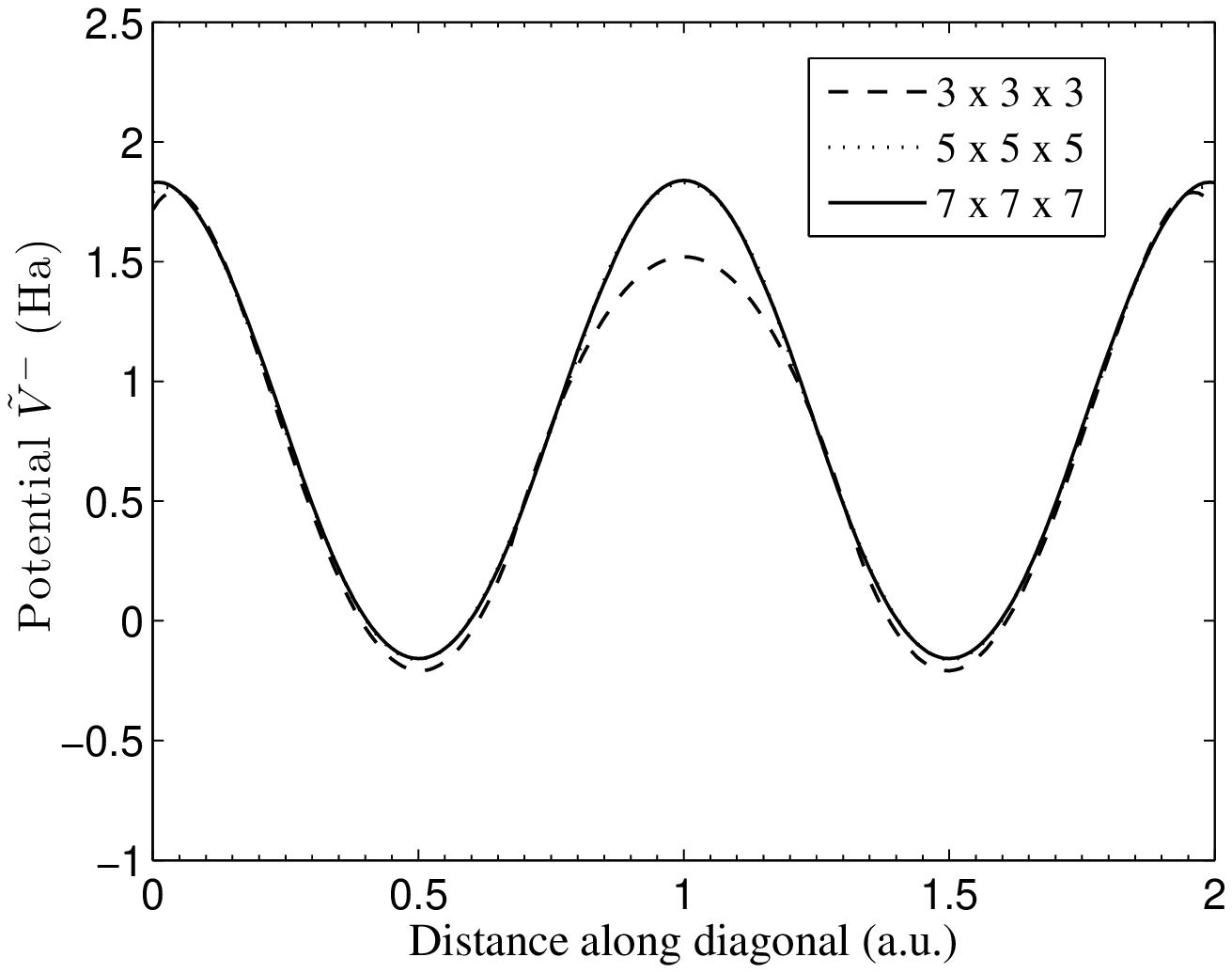, width=0.49\textwidth}}
}
\mbox{
\subfigure[]{\epsfig{file=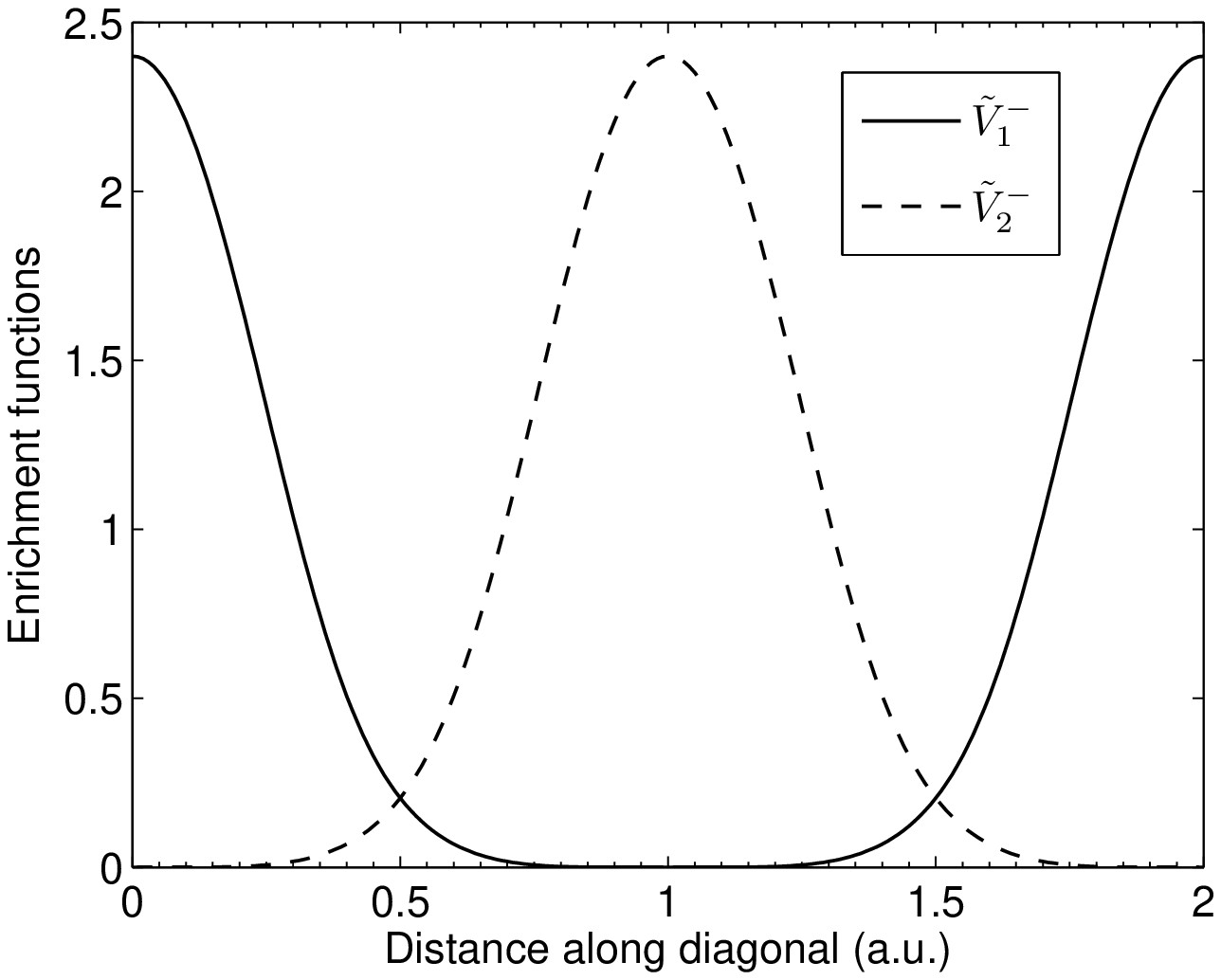, width=0.49\textwidth}}
\subfigure[]{\epsfig{file=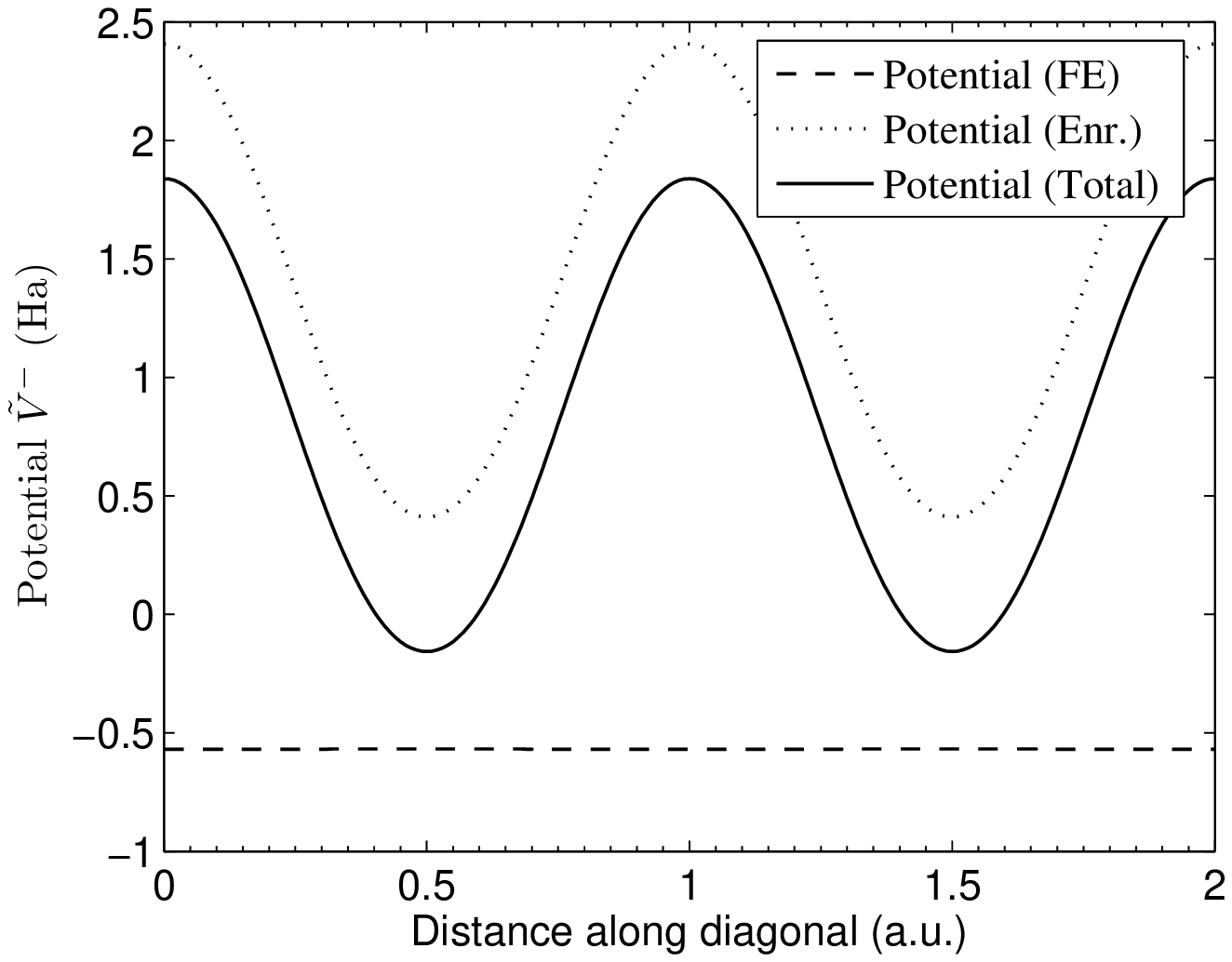, width=0.49\textwidth}}
}

\subfigure[]{\epsfig{file=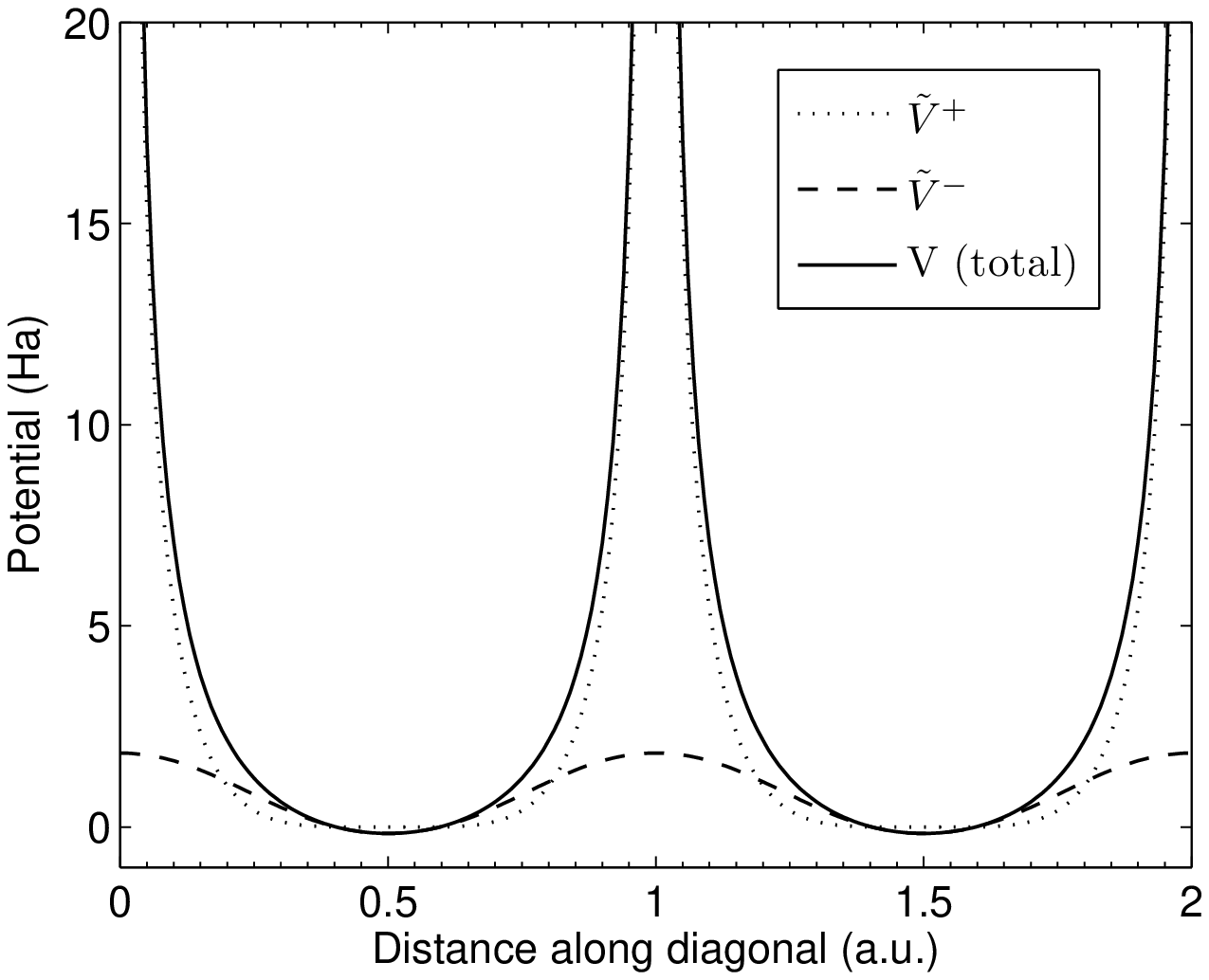, width=0.49\textwidth}}
\caption{Finite element (FE) and enriched finite element (EFE) solutions for the Ewald problem.
(a) Charge densities; (b) FE solutions;
(c) Enrichment functions; (d) EFE solution ($12 \times 12 \times 12$ mesh);
and (e) Total EFE solution ($12 \times 12 \times 12$ mesh).}\label{fig:ewald.1}
\end{figure}
\begin{figure}
\centering
\subfigure[]{\epsfig{file=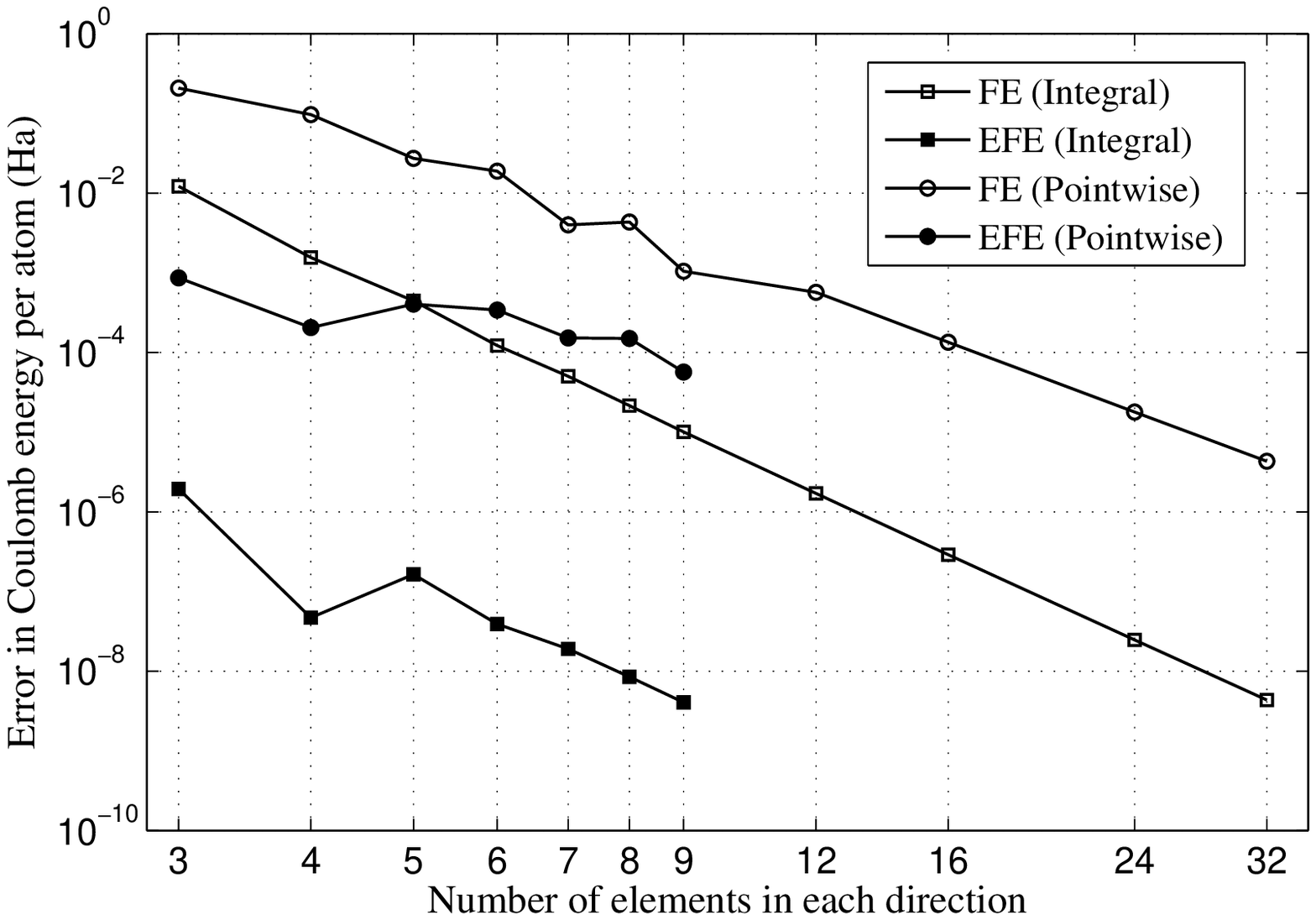, width=0.8\textwidth}}
\subfigure[]{\epsfig{file=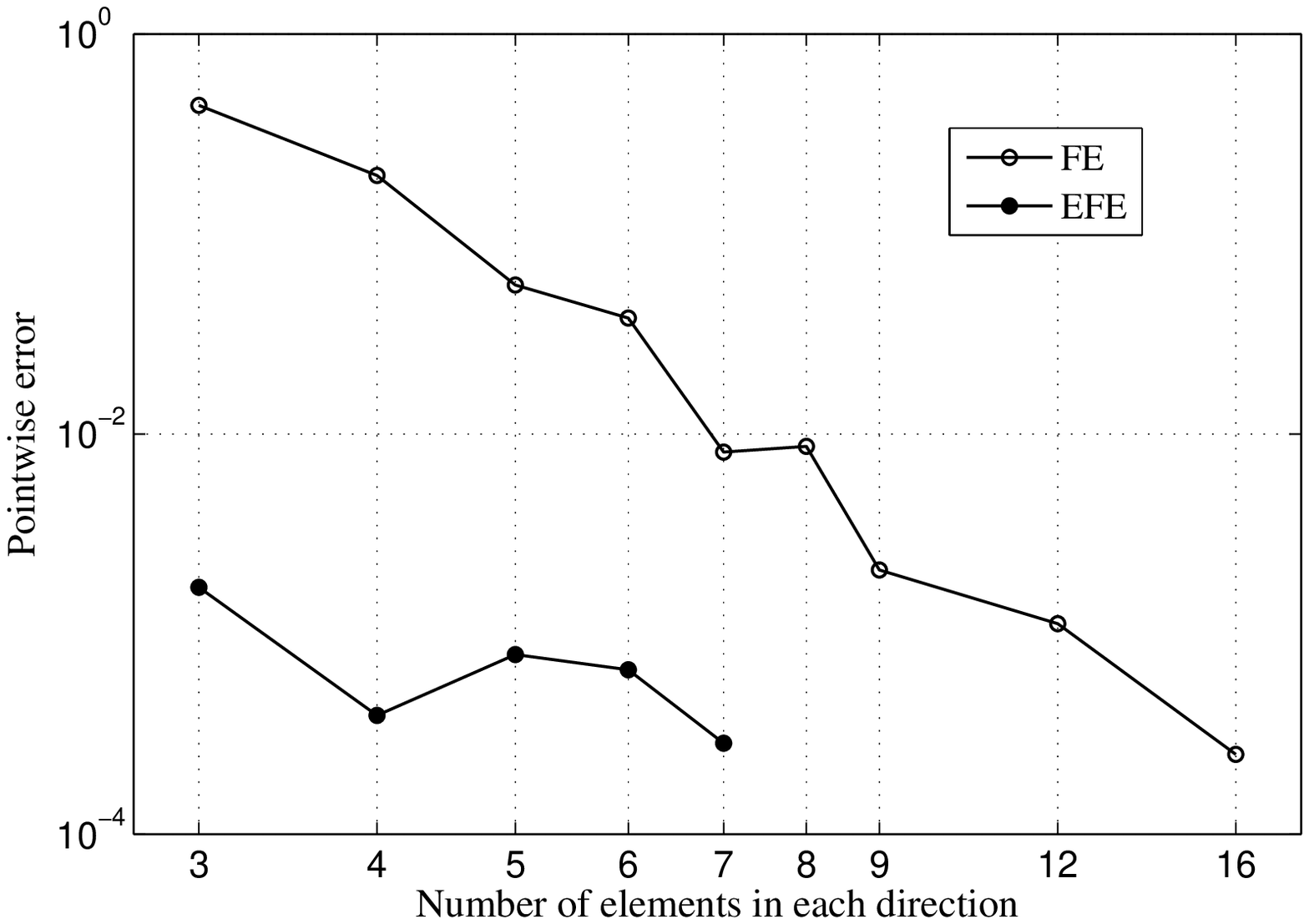, width=0.8\textwidth}}
\caption{Convergence curves. (a) Error in Coulomb energy per atom. Convergence rates: FE (integral) is 6.06,
EFE (integral) is 6.15, FE (pointwise) is 4.97, and EFE (pointwise) is 3.83.
(b) Error in $\sum_{I=1}^2 \tilde{V}^-(\tau_I)$. EFE solution on $16 \times 16 \times 16$ mesh
is used as reference.}
\label{fig:ewald.2}
\end{figure}

The numerical results for the Ewald problem are shown in~\fref{fig:ewald.1}.
The electronic, neutralizing, and neutralized electronic charge densities are
shown in~\fref{fig:ewald.1}a. The FE
potential solution corresponding to the neutralized electronic charge
density is plotted along the diagonal in~\fref{fig:ewald.1}b for different
meshes; convergence is observed as the number of elements along each coordinate
direction increases from 3 to 5 to 7.
The two enrichment functions, one for each atom, are shown
in~\fref{fig:ewald.1}c, and the EFE solution for a $12 \times 12 \times 12$
mesh is depicted in~\fref{fig:ewald.1}d.  The total potential solution,
which includes the singular nuclear potential contributions, 
is shown in~\fref{fig:ewald.1}e.
Figure~\ref{fig:ewald.2}a shows the convergence curves for the Coulomb energy
per atom for FE and EFE solutions. 
Numerical integration in the EFE solution
is carried out with a $20 \times 20 \times 20$ Gauss
quadrature rule in each element, which ensures that the quadrature
error is less than the approximation error. From~\fref{fig:ewald.2}a, we
see that use of the pointwise expression~(\ref{eq21})
adversely affects both accuracy and rate of convergence. To explore this further,
the pointwise error $\sum_{I=1}^2 \tilde{V}^-(\tau_I)$ is plotted
in~\fref{fig:ewald.2}b, where the EFE solution ($16 \times 16 \times 16$ mesh) is used
as reference. Clearly, the pointwise error is
appreciable, which explains the decrease in accuracy and the
non-monotonic convergence in the curves plotted
in~\fref{fig:ewald.2}a. The enriched finite element solution provides an
accuracy of order $10^{-8}$ Ha in the Coulomb energy with 9
elements in each direction whereas with finite elements, 32 elements
in each direction is required for the same accuracy. On using the integral
expression in~\eref{eq66} to compute the Coulomb energy, the 
optimal sextic rate of
convergence is obtained with FE and EFE (\fref{fig:ewald.2}a), consistent with theory~\cite{StrF73}.

\begin{figure}[htb]
\psfrag{rcn = 1/2, rce = 1}[m][c]{$\scriptstyle r_{cn} \, = \, 1/2,\;r_{ce}\,=\,1$}
\psfrag{rcn = 1/3, rce = 1}[m][c]{$\scriptstyle r_{cn} \, = \, 1/3,\;r_{ce}\, = \, 1$}
\psfrag{rcn = 1/3, rce = 2/3}[m][c]{$\!\scriptstyle r_{cn}\, =\, 1/3,\;r_{ce}\, =\, 2/3$}
\centering
\epsfig{file=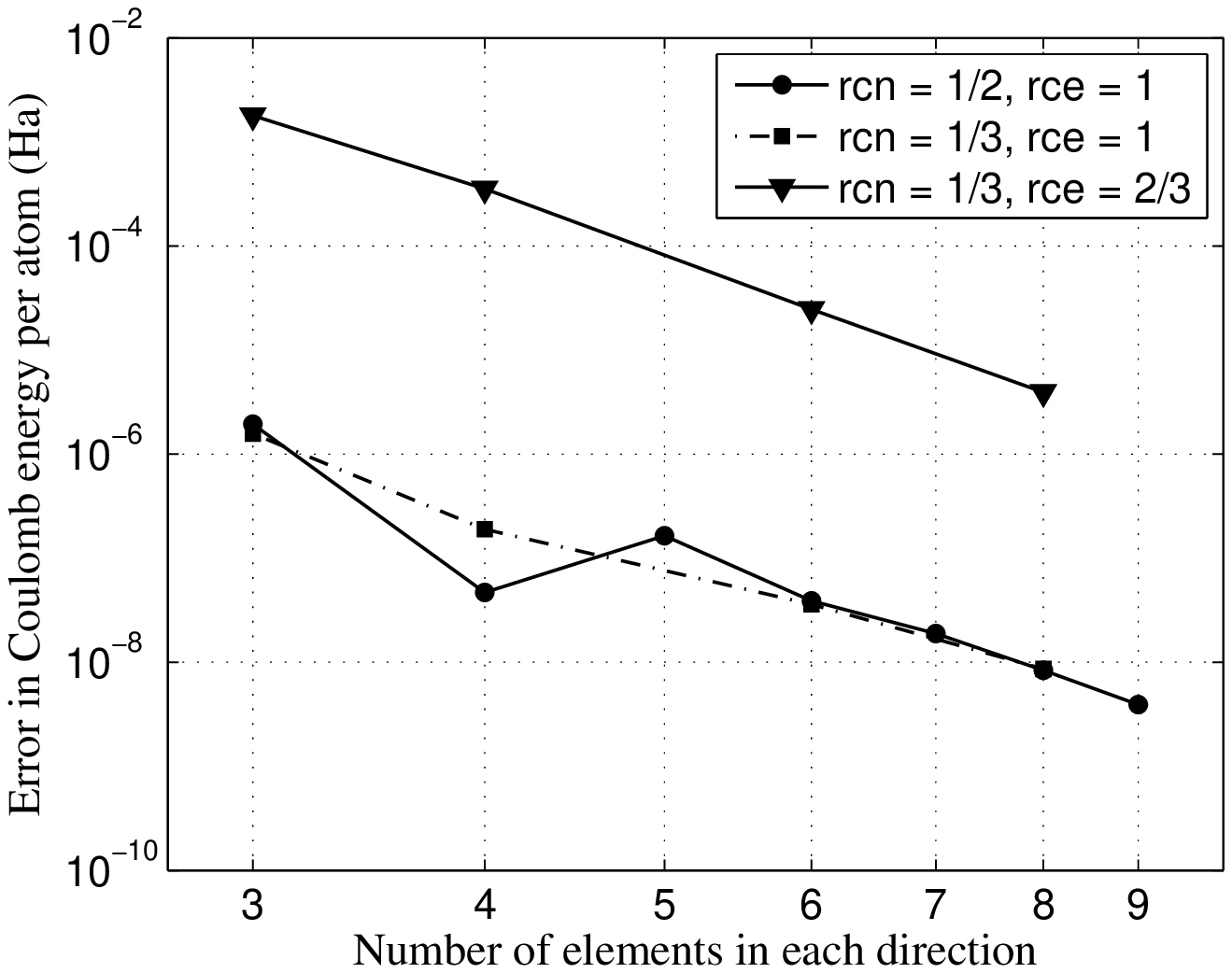,width=0.65\textwidth}
\caption{Error in Coulomb energy per atom for EFE computations using different
$r_{cn}$ and $r_{ce}$.}
\label{fig:ewald.3}
\end{figure}

From the error curve for EFE in~\fref{fig:ewald.2}a (integral
expression), we note that the mesh $4\times4\times4$ delivers
much better accuracy and convergence rate than the other
meshes. Further analysis helps to explain this anomaly. For the
Ewald problem, the corner atom is at the origin and the
center atom is at $\tauv=(1/2, 1/2, 1/2)$ 
(length of the diagonal is 2 a.u.), and $r_{cn}=1/2$ a.u.\ ensures no
overlap between the neutralizing densities of the two atoms.
Furthermore, the support of the neutralizing charges coincides with 
the location of the
nodes ($4 \times 4 \times 4$ mesh) along the diagonal of the cube.  
We repeat the calculations for two more cases: $r_{cn}=1/3$
a.u., $r_{ce} = 1$ a.u.; and $r_{cn}=1/3$ a.u., $r_{ce} = 2/3$ a.u. 
The results obtained for all three cases are shown 
in~\fref{fig:ewald.3}.  One can observe that the solutions for the 
mesh $4\times4\times4$ with $r_{cn}=1/3$ a.u.\ are consistent with 
the solutions on other meshes, and hence the 
markedly better accuracy for the case when $r_{cn}=1/2$ a.u.\
(Figures~\ref{fig:ewald.2}a and~\ref{fig:ewald.3}) 
is a consequence of the choice $r_{cn} = 1/2$ a.u.

\begin{figure}[htb]
\psfrag{rcn = 1/2}[m][c]{$\scriptstyle r_{cn} \, = \, 1/2$}
\psfrag{rcn = 1/3}[m][c]{$\scriptstyle r_{cn} \, = \, 1/3$}
\centering
\epsfig{file=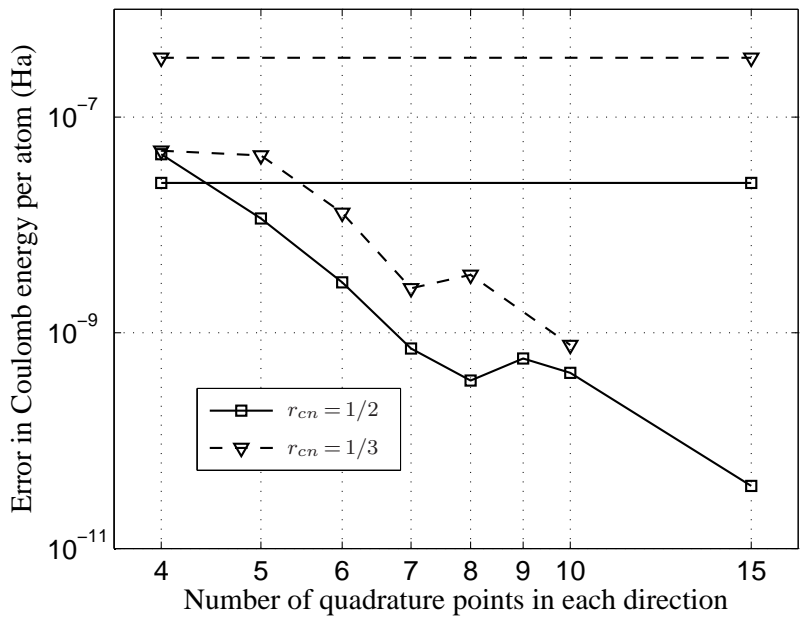, width=0.65\textwidth}
\caption{Comparison of quadrature errors for FE computations on $24 \times 24 \times 24$
mesh. Horizontal lines show the best solution error obtained for each case.}
\label{fig:ewald.4}
\end{figure}

Numerical integration to compute the weak form integrals is carried out using
standard tensor-product Gauss quadrature. Since the
right-hand-side of the Poisson equation~\eref{eq4} is not a polynomial, the
degree of the quadrature rule must be selected so that the integration
error is at least an order smaller than the approximation error in order to obtain variational results.
\fref{fig:ewald.4} shows the error in the Coulomb energy per atom 
($24 \times 24\times 24$ FE mesh) as a function of the number of
quadrature points in each direction.  The error in the
FE solution is shown by the horizontal lines, and~\fref{fig:ewald.4} reveals
that at least five points ($5 \times 5 \times 5$ quadrature
rule) are needed to ensure that quadrature error is below 
approximation error.

\subsection{Diamond}
We now consider diamond, with unit cell defined by lattice vectors
\begin{align*}
\av_1&=a/2(0,1,1), \\
\av_2&=a/2(1,0,1), \\
\av_3&=a/2(1,1,0)
\end{align*}
and carbon atoms at positions
\begin{align*}
\tauv_1&=(0,0,0), \\
\tauv_2&=(1/4,1/4,1/4),
\end{align*}
where $a=6.75$ bohr. For the all-electron problem, we consider a 
total charge density $\rho$ consisting of nuclear point charges 
$q_i=6$ at positions $\tauv_i$ and electronic density $\rhom$ from 
overlapping all-electron atomic densities.

To compute the energy using the formulation (\ref{eq21}) or~(\ref{eq66}), 
we take $g(r)$ as in (\ref{eq16}) with $r_c=1.4$ a.u., so that the 
neutralizing densities $\rhoIt(\xv)= 6 g(|\xv-\tauv_I|)$ are 
nonoverlapping. The total neutralizing density in the cell 
is $\rhot(\xv) = \sum_I \rhoIt(\xv)$, where the sum over positions $I$ 
in the unit and nearest neighboring cells is sufficient. The 
electronic charge density is $\rhom=\sum_I \rhoIm$, which is the sum of all-electron 
atomic densities, where the sum over positions in nearest and second-nearest 
neighbor cells is sufficient to capture the tails of the $\rhoIm$. The 
potentials corresponding to neutralized electronic densities 
$\rhoImt=\rhoIm+\rhoIt$ at each site $I$ are then $\vImt=\vIm+\vIt$, 
where $\vIm$ is the all-electron atomic potential and $\vIt$ is the 
potential corresponding to $\rhoIt$. 
The integral in the second term of (\ref{eq66}) involving
$\rhom(\xv)$ is computed within a sphere of radius $r_c$ using
Gauss quadrature in spherical coordinates. A tensor-product quadrature 
rule with 531 points (spline-fit has 177 knots)
in the radial direction, and 20 points in each of the 
two angular directions is used. We obtain $I_{sph}=-9.44150186$ with
precision to all digits shown.
The enrichment functions $\Valmt$ are formed as in (\ref{eq33}), 
with the $\vImt$ brought smoothly to zero at radius $a/\sqrt{3}$ 
in order to maximize their extents consistent with summation over
 unit and nearest neighbor cells only. The required potential $\Vmt$ 
corresponding to the neutralized electronic density $\rhomt$ is then 
obtained from the enriched FE solution of the associated Poisson 
equation, with enrichment functions $\Valmt$; whereupon the Coulomb 
potential and energy are obtained from (\ref{eq2}) and (\ref{eq21}) 
or (\ref{eq66}), respectively.

\begin{figure}[!hbp]
\centering
\psfrag{rho-}[B][l]{~~~~$\rho^-$}
\psfrag{rho~}[B][l]{~~~~$\tilde{\rho}$}
\psfrag{rho-~}[B][l]{~~~~~~$\tilde{\rho}^-$}
\psfrag{V+~}[B][l]{~~~~~~$\tilde{V}^+$}
\psfrag{V-~}[B][l]{~~~~~~$\tilde{V}^-$}
\psfrag{V (total)}[B][l]{~~~~~~~~~~~~~$V$ (total)}
\mbox{
\subfigure[]{\epsfig{file=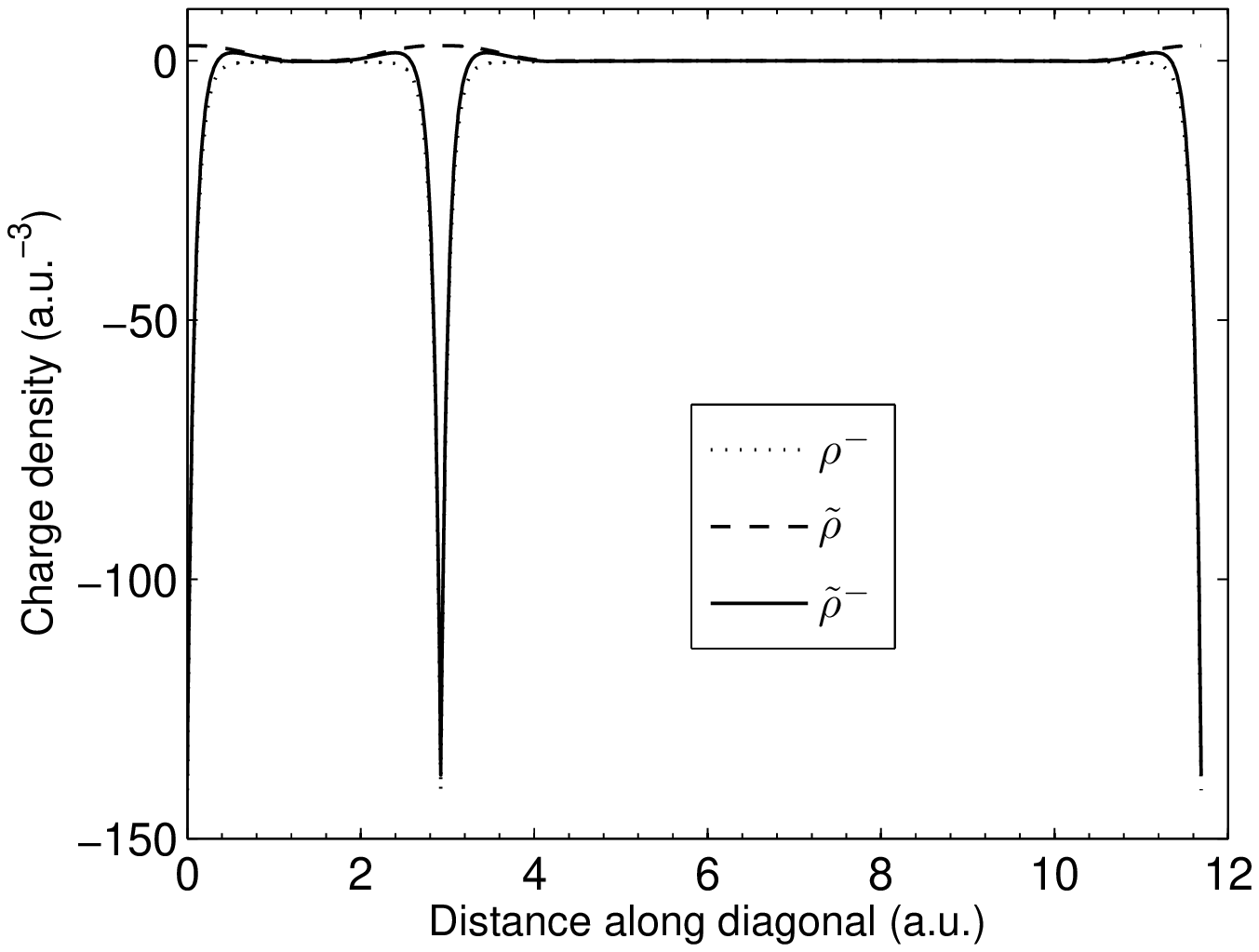, width=0.47\textwidth}}
\subfigure[]{\epsfig{file=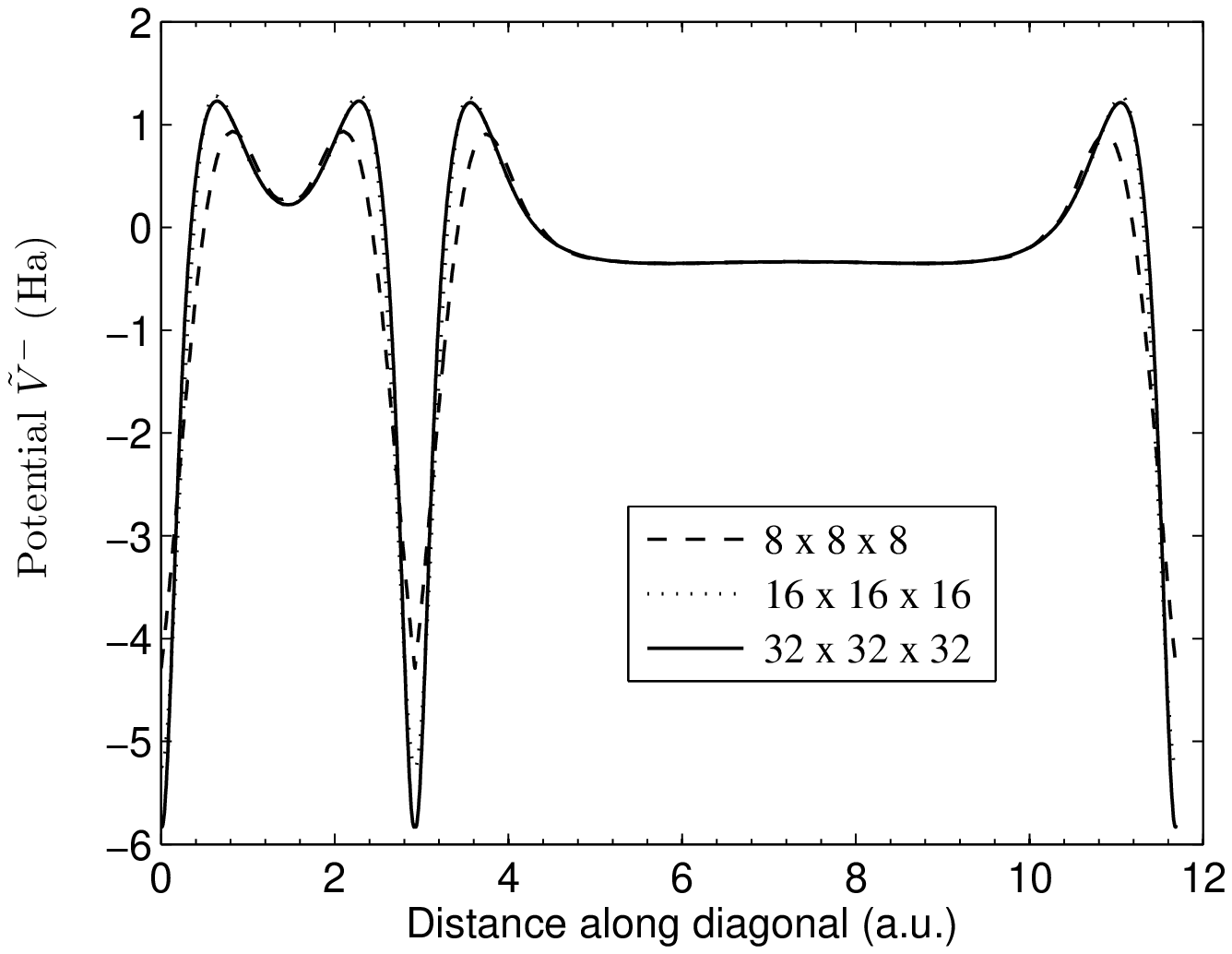, width=0.47\textwidth}}
}
\mbox{
\subfigure[]{\epsfig{file=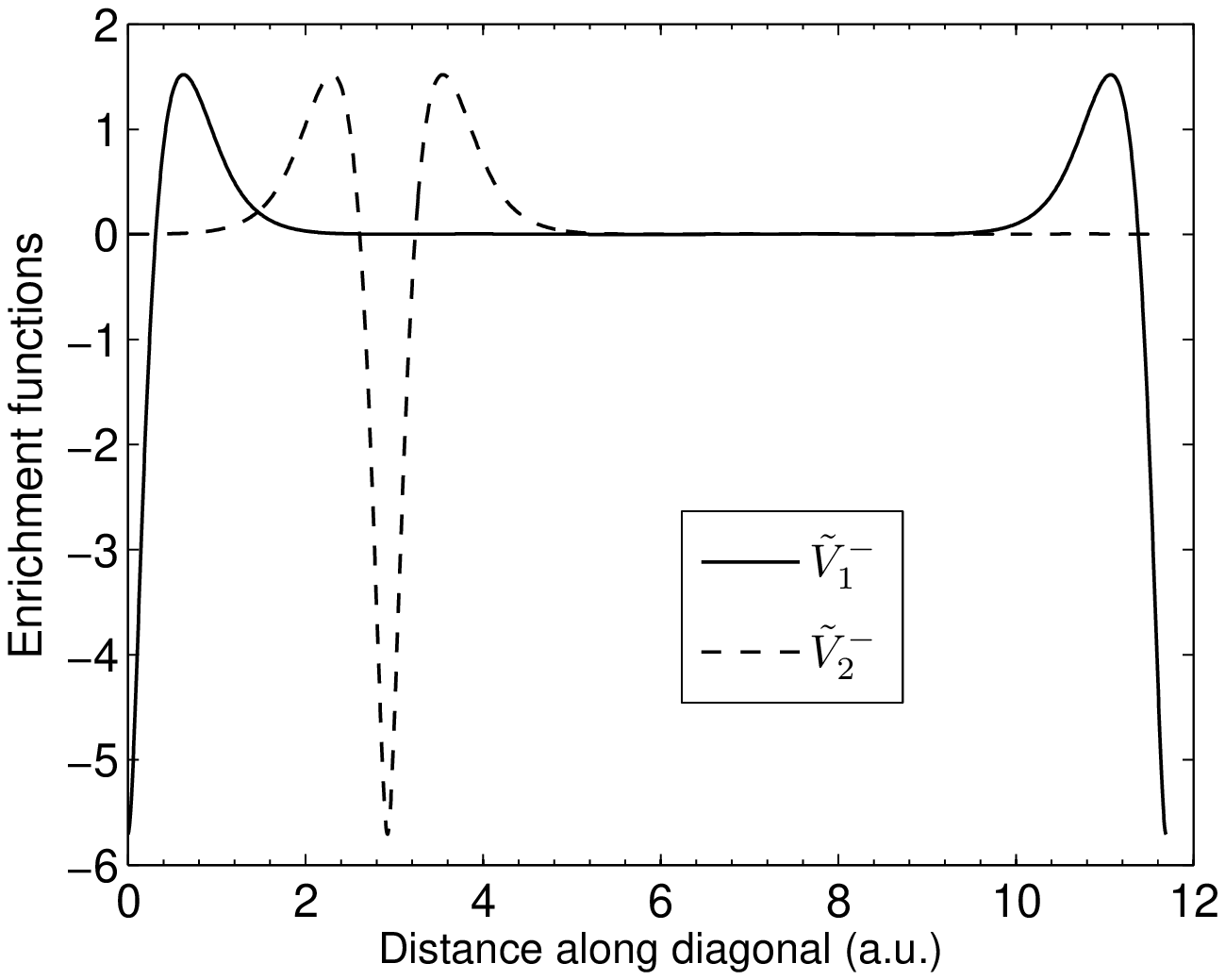, width=0.47\textwidth}}
\subfigure[]{\epsfig{file=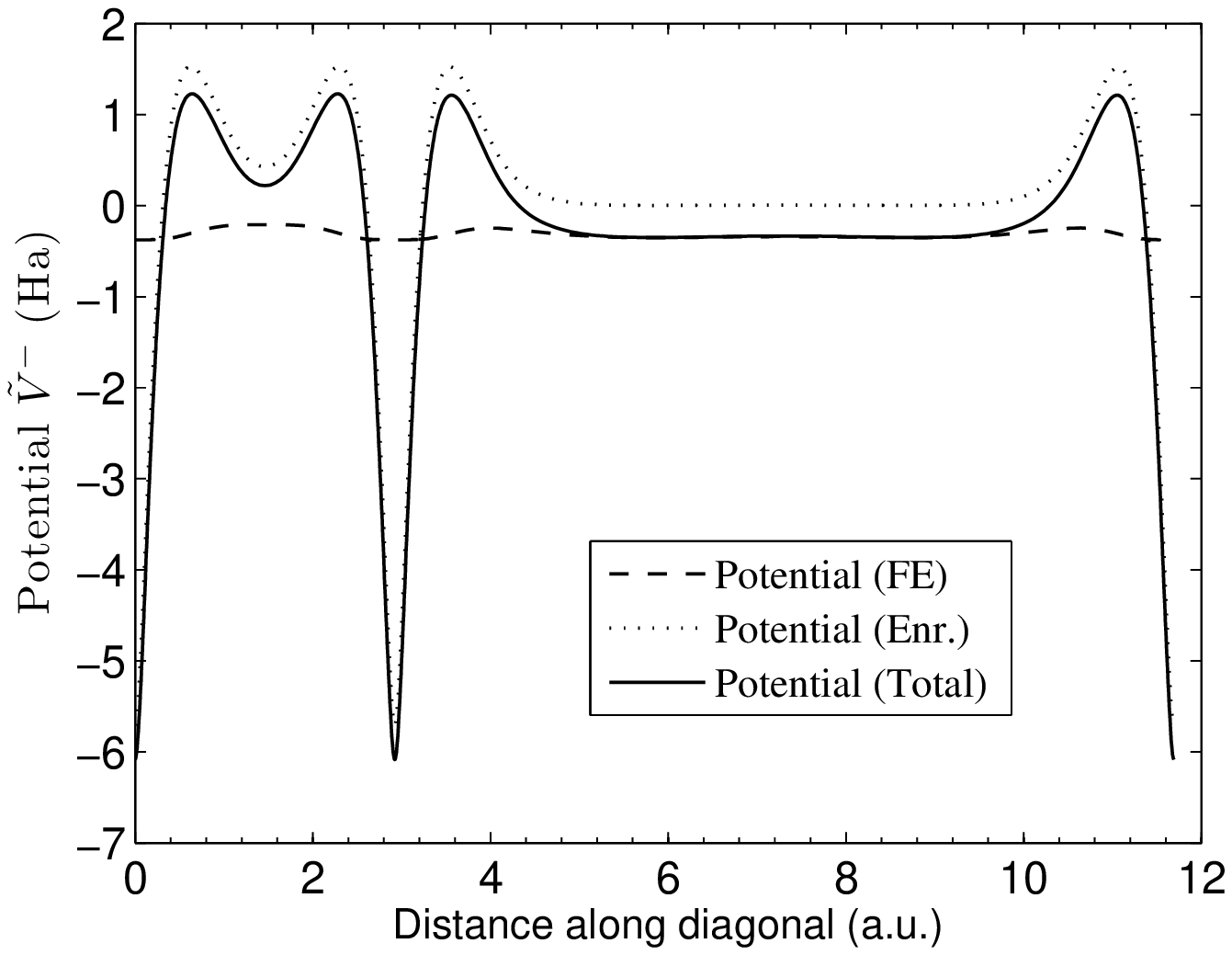, width=0.47\textwidth}}
}
\mbox{
\subfigure[]{\epsfig{file=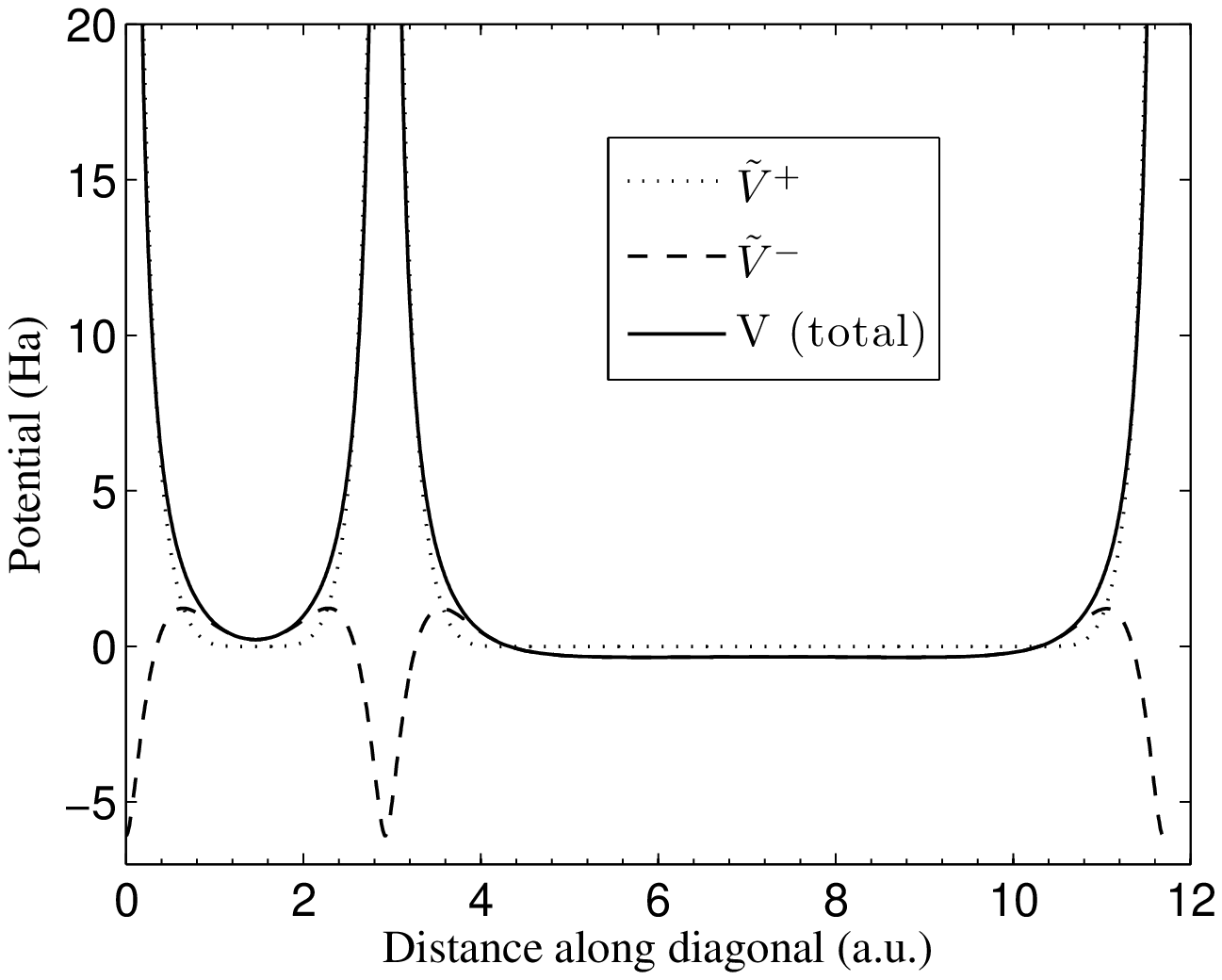, width=0.47\textwidth}}
\subfigure[]{\epsfig{file=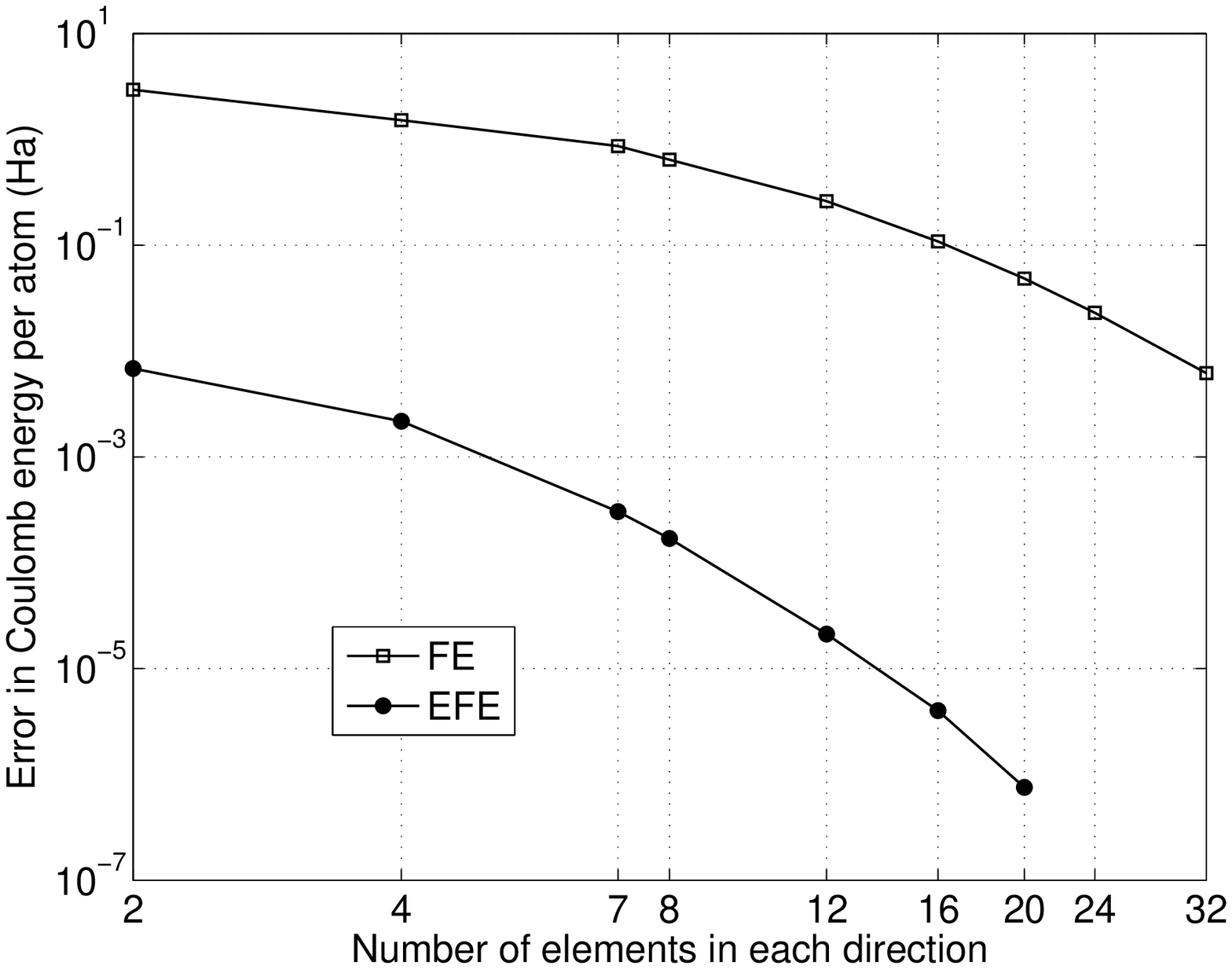, width=0.47\textwidth}}
}
\caption{Finite element (FE) and enriched finite element (EFE) solutions for crystalline
diamond. (a) Charge densities; (b) FE solutions;
(c) Enrichment functions; (d) EFE solution ($24 \times 24 \times 24$ mesh);
(e) Total EFE solution ($24 \times 24 \times 24$ mesh); and
(f) Error in Coulomb energy per atom (integral expression). 
    Convergence rate for EFE: meshes 8-12-16 give 5.38 and meshes 12 and 16 yield 5.79.
    EFE solution on $24 \times 24 \times 24$ mesh is used as reference.}
\label{fig:carbon}
\end{figure}

The electronic, neutralizing, and neutralized electronic charge densities are 
shown in~\fref{fig:carbon}a. The neutralized finite element potential
solutions along the diagonal of the unit cell are plotted 
in~\fref{fig:carbon}b for 8, 16, and 32 elements along each primitive
lattice direction. The two enrichment functions, one for each atom, 
are shown in~\fref{fig:carbon}c, and the 
neutralized EFE potential solution and total potential solution
for the $24 \times 24 \times 24$ mesh are 
illustrated in~Figures~\ref{fig:carbon}d and~\ref{fig:carbon}e.
The error in the Coulomb energy with mesh refinement is 
plotted in~\fref{fig:carbon}f for FE and EFE methods; with EFE result 
on a $24 \times 24 \times 24$ mesh as reference. 
The enriched finite element solution has an accuracy of 
$4 \times 10^{-6}$ Ha in the Coulomb energy
for a $16 \times 16 \times 16$ mesh (28,674 degrees of freedom),
whereas the the best uniform-mesh FE result provides an accuracy of
only $6 \times 10^{-3}$ Ha on a $32 \times 32 \times 32$ mesh (229,376
degrees of freedom). 
While the use of adaptive higher-order finite elements
will require fewer basis functions than uniform FE,
previous studies~\shortcite{GoeI98,TorEE06,BylHW09,LehHP09,phanish:2010:NPF} suggest that their performance in terms of number of basis
functions for a prescribed accuracy may not compare favorably
to EFE since they do not incorporate physics-based
knowledge (nature of variations, in addition to scale) within the approximation space.

The numerical integration for FE and EFE solutions is performed using 
tensor-product Gaussian quadrature rules. For EFE computations, the number of 
integration points are increased until the integration error 
is below the approximation error. Due to the large values and sharp
variations of the enrichment functions in the vicinity of the atomic
positions, higher-order quadrature is required in finite
elements that are close to the atoms.  As a result, we use
higher-order quadrature in elements with at least one vertex
at an atomic position, and a lower-order quadrature rule in all other
elements. The reference solution for the Coulomb energy per atom 
in~\fref{fig:carbon}f is $-75.985203$ Ha, which is obtained using a
$25 \times 25 \times 25$ quadrature rule in elements that are in
the vicinity of the atoms and a $15 \times 15 \times 15$ quadrature rule
in other elements. The integration scheme we adopt is the simplest rule that
provides us with the required accuracy. For better efficiency, 
a possible approach is to use tetrahedral mesh generation techniques
to construct a partition of a finite element
with an atom (Coulomb singularity) located at one of its vertices or
in its interior. With a graded tetrahedral mesh that is focused
towards the atomic position, standard quadrature rules
within the tetrahedral elements would suffice to accurately 
integrate the neutralized charge density and the enrichment functions.
Within an EFE method, the tetrahedral mesh so constructed will
be solely used for the purpose of numerical integration; the number of degrees of
freedom remain unchanged. For the Coulomb singularity,
\citeN{Bat00} and~\shortciteN{havu:2004:NET} adopt the Duffy 
transformation~\cite{duffy:1982:QOP} to numerically integrate
matrix-elements with $1/r$ factors. The development of 
accurate and efficient quadrature schemes
in EFE methods is a topical issue at the forefront of current research; recent work in
this direction includes a generalization of the Duffy transformation for
integrating power singularities~\cite{mousavi:2010:GDT}.
Adaptive integration schemes or quadrature rules that are tailored to
the form of the enriched FE basis functions could prove to be more efficient 
to solve the neutralized Poisson problem. This and related topics
are subjects of ongoing research.


\section{Conclusions}\label{sec5}
We have presented a linear scaling formulation for the solution of the all-electron Coulomb 
problem in crystalline solids. The resulting method includes full nuclear potentials, with no smearing approximations, is systematically 
improvable, and well suited to large-scale quantum mechanical 
calculations in condensed matter in which the Coulomb potential and energy of a 
continuous electronic and singular nuclear density 
are required. Linear scaling is achieved by introducing smooth, 
strictly local neutralizing densities to render nuclear interactions 
strictly local, and solving the remaining neutral Poisson problem 
for the electrons in real space. Rapid variations of the electronic density in the vicinity of the nuclei were efficiently treated using enriched FE methods, with isolated atomic solutions as enrichments. By considering different interaction terms, 
we derived two equivalent expressions for the Coulomb energy per unit cell---one
involving a pointwise evaluation of the neutralized electronic potential, 
the other requiring the evaluation of a spherical integral. By avoiding pointwise evaluation of the $C^0$ FE solution, the integral expression proved superior in both accuracy and convergence rate with respect to the number of elements.
For the Ewald problem, accuracy
of order $10^{-8}$ Ha was realized with enriched FE on a mesh with
5105 degrees of freedom. Comparable accuracy on a uniform FE mesh
required 229,376 degrees of freedom. For the Coulomb energy of diamond, the enriched FE solution with 28,674 degrees of freedom
attained an accuracy of $4 \times 10^{-6}$ Ha.

While the calculations here employed a finite element basis, the 
formulation applies quite generally to any desired basis for the 
residual Poisson solution and so should prove of wide
applicability. For example, in the context of high-accuracy
Gaussian based calculations, where fast multipole methods become more
costly, the present method may provide an attractive alternative for
$O(N)$ parallel solution. In such a case, the strictly local
polynomial neutralizing functions employed here might be replaced by
correspondingly localized Gaussians. Any convenient functional form
satisfying the required non-overlap conditions can be employed to equal effect. 

\section*{Acknowledgments}
This work was performed under the auspices of the U.S. Department of 
Energy by Lawrence Livermore National Laboratory under 
Contract DE-AC52-07NA27344. We gratefully acknowledge support from the
Laboratory Directed Research and Development Program; the National Science
Foundation through contract grant DMS-0811025 to the
University of California at Davis; and additional financial support from
the UC Lab Fees Research Program.

\bibliographystyle{chicago} 
\bibliography{JEP-IJMCE,poisson}

\end{document}